\documentclass[aps,prb,twocolumn,floatfix]{revtex4-1}
\usepackage{graphicx}
\usepackage{amsmath}
\usepackage{color}
\usepackage{comment}
\usepackage{bm}

\newcommand{\addMR}[1]{\textcolor{red}{#1}}

\newcommand{\Ket}[1]{\vert  #1 \rangle}

\newcommand{\Avg}[1]{\langle  #1  \rangle}

\newcommand{\be}{\begin{equation}}
\newcommand{\ee}{\end{equation}}
\newcommand{\bea}{\begin{eqnarray}}
\newcommand{\eea}{\end{eqnarray}}

\newcommand{\bpm}{\begin{pmatrix}}
\newcommand{\epm}{\end{pmatrix}}

\renewcommand{\phi}{\varphi}
\renewcommand{\epsilon}{\varepsilon}
\renewcommand{\vec}[1]{{\bf #1}}

\newcommand{\<}{\langle}
\renewcommand{\>}{\rangle}

\newcommand{\bk}{\textbf{k}}
\newcommand{\bd}{\textbf{d}}
\newcommand{\hbd}{\hat{\textbf{d}}}

\newcommand{\hbz}{\hat{\textbf{z}}}

\begin{document}

\title{Anomalous edge states and the bulk-edge correspondence for\\ periodically-driven two dimensional systems}

\author{Mark S. Rudner$^{1,2,3}$, Netanel H. Lindner$^{3,4}$, Erez Berg$^{2,5}$, and Michael Levin$^6$}
\affiliation{ 
$^{1}$ The Niels Bohr International Academy, Niels Bohr Institute, Blegdamsvej 17, DK-2100 Copenhagen, Denmark\\
$^{2}$ Department of Physics, Harvard University, Cambridge, MA 02138, USA\\
$^3$ Institute of Quantum Information and Matter, California Institute of Technology, Pasadena, CA 91125, USA\\
$^4$ Department of Physics, California Institute of Technology, Pasadena, CA 91125, USA\\
$^5$ Department of Condensed Matter Physics, Weizmann Institute of Science, Rehovot 76100, Israel\\
$^{6}$ Condensed Matter Theory Center, Department of Physics, University of Maryland, College Park, Maryland 20742, USA
}

\begin{abstract}
Recently, several authors have investigated topological phenomena in periodically-driven systems of non-interacting particles.
These phenomena are identified through analogies between the Floquet spectra of driven systems and the band structures of static Hamiltonians.
Intriguingly, these works have revealed phenomena which cannot be characterized by analogy to the topological classification framework for static systems. 
In particular, in driven systems in two dimensions (2D), robust chiral edge states can appear even though the Chern numbers of all the bulk Floquet bands are zero.
Here we elucidate the crucial distinctions between static and driven 2D systems, and construct a new topological invariant that yields the correct edge state structure in the driven case.
We provide formulations in both the time and frequency domains, which afford additional insight into the origins of the ``anomalous'' spectra which arise in driven systems.
Possibilities for realizing these phenomena in solid state and cold atomic systems are discussed.
\end{abstract}

\maketitle

The discovery of the quantized Hall effect\cite{integerquantumhall} revealed the existence of
a powerful new class of extremely robust quantum phenomena which can be observed with high
fidelity, largely independent of sample size, shape, and composition, up to macroscopic
(millimeter) scales. The robust nature of these phenomena can be linked to the presence of
energy gaps for bulk excitations, combined with the existence of nontrivial topological
structures associated with the systems' ground state wave functions\cite{TKNN}.
Recently, these ideas were used to predict the existence of new classes of
materials\cite{KaneMele05,Bernevig06}, the topological insulators, which were found
experimentally\cite{Koenig07,Hsieh08} shortly thereafter.

Meanwhile, a wide variety of new experimental tools has been developed for actively controlling and probing the behavior of electronic, cold-atomic, and purely photonic systems.
For cold atoms, various methods have been proposed for creating ``synthetic gauge fields,'' which mimic the effects of magnetic fields\cite{FetterRMP,DalibardRMP, Cooper2011} or spin-orbit coupling\cite{Campbell2011,Lin2011} for neutral atoms.
Several groups have also suggested the possibility of using microwave and optical techniques in solids to realize topologically non-trivial effective band structures in ``trivial'' materials \cite{Nate11,Kitagawa11, Oka09, Inoue10, Fertig11, Nate3D, Morell12, Podolsky2013}.
Topological phenomena have been identified for strongly driven\cite{Yao2007, KitagawaPRA, KitagawaPRB, Jiang11, Dahlhaus11, Levchenko12, Dora12, Cayssol12, Reynoso2012} or dissipative quantum systems\cite{Rudner09, Diehl11} as well.
Analogues of topological phenomena in driven systems have even recently been observed in photonic experiments\cite{QWExpt, Segev12}.
These advances motivate the detailed study of topological phenomena in periodically-driven systems.

Here we focus our attention on topological features of the single-particle properties of two-dimensional translationally-invariant tight binding systems.
In the static case, where the Hamiltonian is constant in time, the topological properties of these systems are well understood.
When no additional symmetries are present, a complete topological characterization 
is provided by the set of values of an integer topological invariant, the Chern number, 
evaluated for each band\cite{TKNN, Schnyder08, Kitaev09}.

One of the most striking applications of this topological characterization 
is to the edge physics of these systems.
In a finite geometry with an edge, 
a two-dimensional system may support chiral edge modes which propagate at energies within a bulk band gap.
Remarkably, the values of the Chern numbers can be used to predict the net number of chiral edge modes traversing each bulk gap (counted according to their chirality).
This edge state count is significant because it is a robust property which is independent of the fine details of the system and of the edge.
In the case of the quantized Hall effect, the guaranteed existence of these modes provides a powerful 
framework for explaining a variety of complex phenomena\cite{Halperin82}.

Given the power of this approach, we seek to generalize these results to periodically driven systems.
In the driven case, the analogue of the energy spectrum is the Floquet spectrum -- the set of eigenvalues of the time evolution operator, evaluated over one complete cycle of driving.
The eigenvalues of the unitary evolution operator are unit modulus complex numbers, which are defined on a {\it circle}. 
We express them in terms of a ``quasi-energy'' $\epsilon$ as $e^{-i \epsilon T}$, where 
$T$ is the driving period. 
The spectrum is thus $2\pi/T$-periodic in $\epsilon$.

There are many close analogies between 
static and periodically-driven systems.
In particular, the Floquet spectrum can be organized into quasi-energy bands with associated Chern numbers, just as in the static case.
Also, driven systems may exhibit chiral edge modes when defined in a finite geometry with a boundary.
Despite these similarities, however, the Chern numbers employed in the static case do \emph{not}
give a full characterization of the topological properties of periodically driven systems.
In particular, for driven systems, these invariants do not uniquely determine the number of chiral edge
modes within each bulk band gap.

\begin{figure}[t]
\includegraphics[width=0.95\columnwidth]{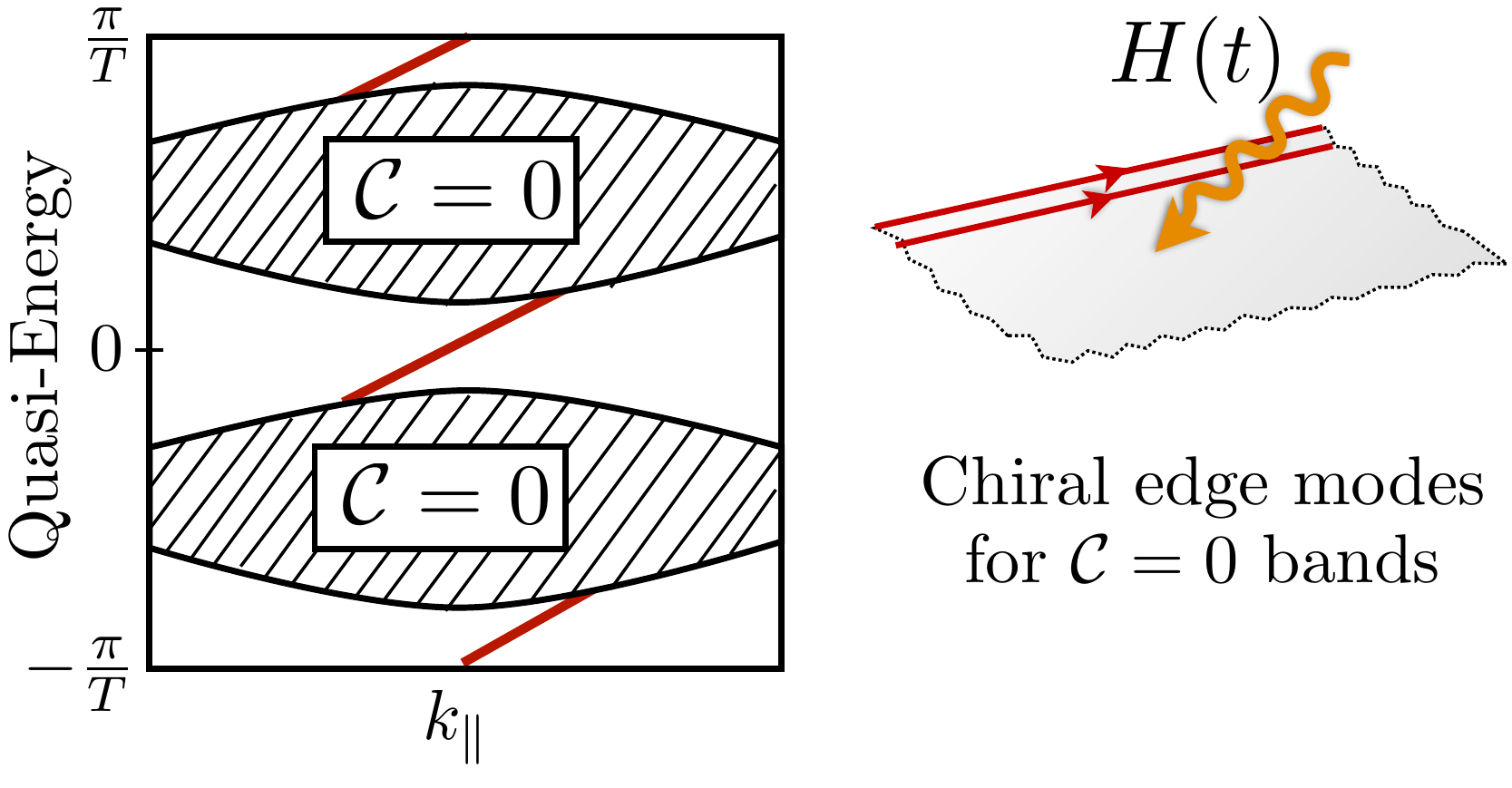}
 \caption[]{Topologically protected Floquet edge modes in a two-band system with trivial Chern indices, $\mathcal{C} = 0$. 
This situation is made possible due to the periodicity of the quasi-energy $\varepsilon$: 
in addition to the edge modes which cross the gap near $\varepsilon = 0$, a second branch of edge modes crosses through
$\varepsilon = \pi/T$, thus connecting the top of the upper Floquet band to the bottom of the lower Floquet band and closing the quasi-energy cycle.
The Chern numbers correctly yield the {\it differences} between the numbers of chiral modes above and below every band, but cannot be used to uniquely determine the edge state spectrum of a periodically-driven system.
}
\label{fig1}
\end{figure}

Recently, new types of edge modes, which cannot be accounted for using
the invariants developed for static systems, were discovered in the Floquet spectra of one and two dimensional periodically-driven systems\cite{KitagawaPRB,Jiang11,QWExpt}.
A schematic picture showing an example of such ``anomalous'' edge modes in a two-band two-dimensional system is shown in Fig.~\ref{fig1}.
Here, two sets of co-propagating chiral edge modes are found, despite the fact that the
Chern numbers associated with both bands are zero. This situation
cannot occur in 
static systems, where such edge modes are only produced at the
boundaries of systems characterized by non-zero Chern numbers.

In this paper, we develop a more complete understanding of this behavior, and identify appropriate topological invariants (winding numbers) for characterizing these new phenomena.
In contrast to the Floquet-band Chern numbers, which only depend on the evolution operator evaluated over one complete driving cycle, the winding numbers 
utilize the information in the evolution operator for {\it all times} within a single driving period.
We show that the winding numbers fully determine the chiral edge mode counts in each of the Floquet gaps.  
Thus our construction provides a complete ``bulk-edge correspondence'' for periodically-driven, two dimensional, single particle systems.

The paper is organized as follows. Sec.~\ref{sec:Failure} gives a
general discussion of the bulk edge correspondence 
for both static and periodically-driven systems. 
In Sec.~\ref{sec:Square} we introduce a simple model of a strongly driven system, similar to the one considered in Ref.~\onlinecite{KitagawaPRB}, which supports a Floquet band structure 
like the one depicted in Fig.\ref{fig1}.
Then in Sec.~\ref{sec:Analysis} we provide a general analysis and construct the winding numbers of the time-dependent evolution operator, 
which fully characterize the topological features of the system. 
The relationship between the winding numbers and the Chern numbers of the Floquet bands is discussed in Sec.~\ref{sec:Analysis}C.
In Sec.~\ref{sec:Nate} we give a complementary approach for deriving the edge state spectrum, which utilizes an analysis in the frequency domain.
We demonstrate this approach for the case of weak, harmonic driving, 
as considered in Refs.~\onlinecite{Nate11,Kitagawa11, Oka09, Fertig11, Nate3D}.
Finally, in Sec.~\ref{sec:Concl} we discuss possible experimental realizations of these new topological spectra, and prospects for future investigation.
Technical aspects of the derivations are provided in the appendix.

\section{Bulk-edge correspondence in static and driven systems}\label{sec:Failure}

One of the main goals of this paper is to explain how the standard correspondence between
the edge state spectrum and the values of bulk topological invariants is modified for
periodically driven systems. 
Specifically, our goal is to identify topological invariants for bulk systems (i.e.~for systems without edges) which give the numbers of protected 
modes that appear in geometries with edges.
Toward this aim, we begin this section with a brief review of
Floquet band theory in order to define what we mean by the ``band structure'' of a periodically driven system.
We then compare and contrast the static and driven cases, and give a concrete example that
demonstrates the situation where intuition from the static case breaks down.

In general, the evolution of a system governed by a time-dependent
Hamiltonian $H(t)$ may be quite complicated. However, in the case
where the Hamiltonian depends {\it periodically} on time, $H(t +
T) = H(t)$ for some driving period $T$, Floquet theory provides a
powerful framework for analysis. In analogy with the usual
expansion in terms of stationary eigenstates of a static
Hamiltonian, here the evolution is conveniently described in terms
of a basis of {\it Floquet states.} These states are solutions to
the time-dependent Schr\"{o}dinger equation of the form
$\Ket{\psi(t)} = \Ket{\Phi(t)}e^{-i\epsilon t}$, where
$\Ket{\Phi(t + T)} = \Ket{\Phi(t)}$. Under the action of the
evolution operator $U(T)$ over one complete period of driving,
each Floquet state is mapped onto itself up to a phase:
$\Ket{\psi(T)} = U(T)\Ket{\psi(0)} = e^{-i\epsilon
T}\Ket{\psi(0)}$. In a stroboscopic sense, the Floquet states play
the role of stationary states for $U(T)$. The parameter
$\epsilon$, called the quasi-energy, is uniquely defined up to
integer multiples of $\omega = 2\pi/T$: any given solution with
quasi-energy $\epsilon$ can also be associated with a quasi-energy
$\tilde{\epsilon} = \epsilon + p\omega$, where $p$ is an integer,
through the relation $\Ket{\tilde{\Phi}(t)} = e^{ip\omega
t}\Ket{\Phi(t)}$. Similar to the crystal momentum of a system with
discrete translational symmetry, the quasi-energy can be thought
of as a periodic variable defined on a quasi-energy Brillouin zone
$-\pi/T < \epsilon \le \pi/T$. When a system has both discrete
time and spatial translation symmetries, the Floquet states are
labeled by $\epsilon$ and the crystal momentum $\vec{k}$. The
Floquet spectrum then consists of a set of bands
$\{\epsilon_n(\vec{k})\}$, where $n$ is a band index.

We now explore the ways in which the topological properties of
Floquet bands are similar to, and different from, those of the
conventional bands of static systems. Consider first a static
(non-driven) two-band system, such as the Haldane
model\cite{haldane}, which exhibits a well-defined band gap. A
Chern number can be defined for each band $n$ by integrating the
Berry curvature associated with the two component spinor
eigenstates $\{\Ket{u_{n}(\vec{k})}\}$ over the Brillouin zone
\cite{TKNN}. If the parameters are such that the Chern numbers of the two
bands are non-zero, we will find chiral edge modes traversing the
band gap when the system is defined in a finite geometry with an
edge. The existence of these edge modes is ``protected,'' in the sense that the modes
cannot be destroyed by any continuous changes of parameters unless the band gap closes.

Mathematically, the Chern number of a band is equal to the {\it difference} between the numbers of chiral edge
modes entering the band from below and exiting above. Because the spectrum of a static system is bounded, there
can never be any chiral edge modes extending beyond the bottom of the lowest band, or above the top of the uppermost band.
Therefore, in a static two-band system, the Chern number uniquely defines the net number of chiral edge states
crossing the band gap.

Now consider a periodically driven system. We can define a Chern
number for each Floquet band $n$ in the same way as in the static
case: the Chern number is obtained by integrating the Berry
curvature associated with the Floquet states
$\{\Ket{\psi_{n}(\vec{k}, t)}\}$, evaluated at any fixed time $t$,
over the crystal momentum Brillouin zone (see Sec.~\ref{sec: relation} for a precise definition). For concreteness we
typically evaluate the Chern numbers using the Floquet state wave
functions at time $t = 0$. However, the Chern number is in fact
independent of the time at which the Floquet states are evaluated
\cite{Nate11,KitagawaPRB}.

Like the static case, the Chern number of a Floquet band is equal to the difference between the numbers
of chiral modes above and below a given Floquet band.
However, due to the periodicity of quasi-energy, the Floquet spectrum is {\it not bounded}.
In particular, in a two-band system, a chiral edge mode that extends out above the top of the ``upper'' band can pass
through the quasi-energy Brillouin zone edge at $\epsilon = \pi/T$, and enter the bottom of the ``lower'' band, from below.
Using the rule that the Chern number gives the difference between the numbers of edge states above and below a band, we
see that chiral edge states {\it can} be found between the two bands of a system with zero Chern numbers, provided that
a ``winding'' edge state also connects them through the quasi-energy zone edge.

To understand how such a situation can arise in a
periodically-driven system, it is useful to consider the following
thought experiment. Suppose we start with a system in which the Floquet bands
have vanishing Chern numbers, and in which there are \emph{no} chiral edge modes
in a finite geometry. The band structure of a
strip, as a function of the momentum $k_\parallel$ along the strip, will resemble the one shown schematically in
Fig.~\ref{figPhaseDiagram}a. Note that, as discussed above, there are
two gaps to consider between the two bands: the gap centered at
zero quasi-energy and the gap centered at $\varepsilon=\pi/T$. Now imagine
that we tune some parameters in the Hamiltonian. As
parameters are changed, one of these gaps, e.g.~the one at
$\varepsilon=\pi/T$, may close and reopen, in such a way that the Chern
numbers of both bands become non-zero. After the reopening of the
gap, the quasi-energy spectrum will look like the one in
Fig.~\ref{figPhaseDiagram}b, with chiral edge modes crossing the
$\varepsilon=\pi/T$ gap. As parameters are varied further, the
$\varepsilon=0$ gap may close and reopen, bringing the Chern
numbers of both bands back to zero. However, the chiral edge modes
in the $\varepsilon=\pi/T$ gap cannot disappear during this
process, since the $\varepsilon=\pi/T$ gap remains open throughout
it. Therefore, after reopening the $\varepsilon=0$ gap, another
chiral edge mode must appear around $\varepsilon=0$
(Fig.~\ref{figPhaseDiagram}c) such that the difference of the number of
chiral edge modes below and above each band is zero. Thus we may
obtain the situation where chiral edge modes exist despite
all Chern numbers being zero.


The example above shows that the {\it bulk} Floquet operator, 
$U(T)$, does not carry sufficient information to predict the number of chiral Floquet edge modes (note that the edge state spectrum will of course be exhibited explicitly if $U(T)$ is evaluated directly for a system with boundaries). 
Here, we construct new invariants defined in terms of the {\it full time-dependent} bulk evolution operator $U(t)$, evaluated for all intermediate times within the driving period.
These invariants contain the missing information needed to predict the complete Floquet edge state spectrum.

\section{Model of anomalous edge states\label{sec:Square}}
\begin{figure}[t]
\includegraphics[width=1.0\columnwidth]{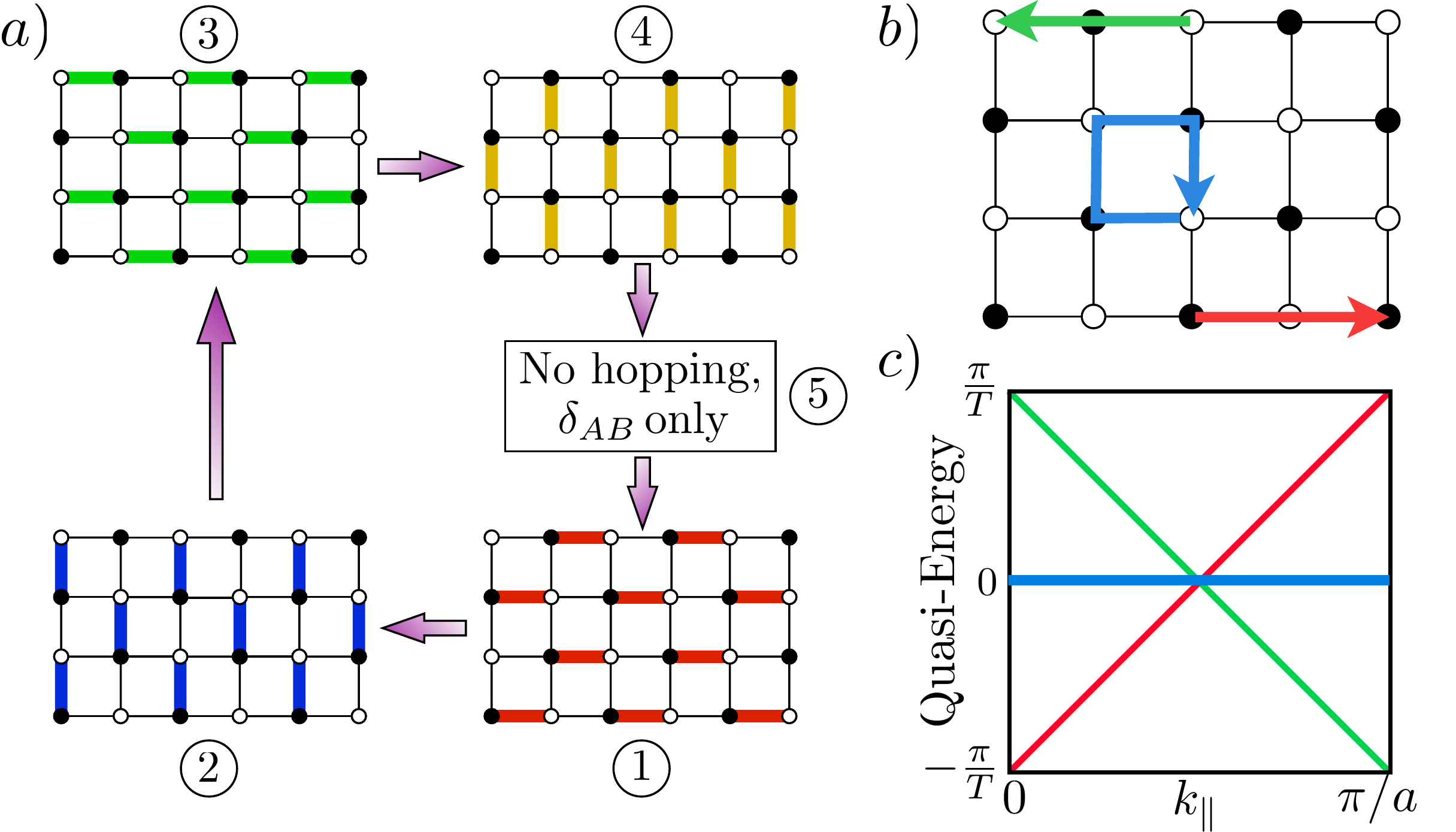}
 \caption[]{Two-dimensional tight-binding model exhibiting anomalous Floquet edge modes. 
a) Driving protocol. Hopping amplitudes are varied in a spatially-homogeneous but chiral, time-periodic way. 
A sublattice potential $\delta_{AB}$ differentiating the two types of sites in the bipartite lattice (filled and open circles) is applied throughout.
During four of the five equal-length phases of the cycle, hopping 
along one of the four distinct bond types is allowed, with amplitude $J$ (bold lines).
During the fifth phase, all hopping amplitudes are zero, while the sublattice potential remains constant.
b) In the simple case $JT = 5\pi/2$, $\delta_{AB} = 0$, particles in the bulk move in closed trajectories encircling plaquettes, and the Floquet operator is the identity.
Chiral edge modes propagate along the boundaries.
c) Floquet spectrum for the case described in b).
}
\label{figSquare}
\end{figure}

In this section we illustrate the breakdown of the traditional bulk-edge correspondence for a Floquet system by analyzing an
explicit two-dimensional tight-binding model.
This model is chosen for conceptual simplicity; a more realistic model for experimental implementations will be discussed in Sec.~\ref{sec:Nate}.
Consider a tight-binding model on a bipartite square lattice, with hopping amplitudes varied in a spatially homogeneous but time-periodic way as shown in Fig.\ref{figSquare}a.
In addition to the cyclic modulation of hopping amplitudes, a constant sublattice potential $\delta_{AB} = \epsilon_A  - \epsilon_B$ which distinguishes between $A$ and $B$ sites (filled and empty circles, respectively) may be applied.
The system's evolution is generated by the time-dependent Hamiltonian 
\begin{eqnarray}
H(t) &=& \sum_{\vec{k}} \left( \begin{array}{cc} c^{\dagger}_{\vec{k}, A} & c^{\dagger}_{\vec{k},B} \end{array} \right)
 H(\vec{k}, t)
 \left( \begin{array}{c} c_{\vec{k}, A} \\ c_{\vec{k},B} \end{array} \right)  \label{Hexample} \\
 H(\vec{k}, t) &=& -  \sum_{n=1}^4 J_{n}(t) \left(e^{i\vec{b}_{n}\cdot \vec{k}}\sigma^+ + e^{-i\vec{b}_{n}\cdot \vec{k}}\sigma^-\right)
+ \delta_{AB}\sigma_z. \nonumber
\end{eqnarray}
Here $c^{\dagger}_{\vec{k}, \alpha}$ 
creates a particle in a Bloch state with crystal momentum $\vec{k}$ on sublattice $\alpha = \{A, B\}$, 
and $J_{n}$ controls hopping from each $B$ site to its neighboring $A$ site along the highlighted bond in step $n$, shown in Fig.\ref{figSquare}a. 
The Pauli matrices $\sigma_z$ and $\sigma^{\pm} = (\sigma_x \pm
i\sigma_y)/2$ act in the sublattice space. The vectors
$\{\vec{b}_i\}$ are given by $\vec{b}_1 = -\vec{b}_3 = (a,0)$ and
$\vec{b}_2 = -\vec{b}_4 = (0,a)$.

One driving cycle consists of five equal length segments, of duration $T/5$, where $T$ is the driving period.
During the $n$-th segment of the cycle, where $n = 1,\ldots,4$, the hopping amplitude $J_n$ is set to a value $J$, while the other three hopping amplitudes are set to 0.
During step 5 all hopping amplitudes are set to 0, but the sublattice potential is still allowed to act.
This ``holding period'' is needed in order to ensure that the model has a sufficiently rich phase diagram to fully illustrate the topological classification that we will develop below.
The driving protocol is inherently chiral, as the cycle can be executed in two inequivalent patterns ($1-2-3-\cdots$ or $5-4-3-\cdots$).

For an infinite system (or a finite system with periodic boundary conditions), translational invariance can be exploited to reduce the problem of finding the Floquet operator for this system to multiplying a small number of Pauli matrices.
The simplest case, illustrated in Figs.\ref{figSquare}b and c, occurs when $JT/5 = \pi/2$ and $\delta_{AB} = 0$.
Here, a particle moves with probability 1 between neighboring sites during each hopping step of the cycle.
As shown by the light blue trajectory in Fig.\ref{figSquare}b, over one complete driving cycle each particle makes a loop around a plaquette and returns to its initial position.
Therefore the bulk Floquet operator for this case is simply the identity.
The Floquet spectrum features two degenerate bands collapsed at quasi-energy zero.
Because the bands are fully collapsed, none of the standard invariants for two-dimensional systems can take non-trivial values.

What happens in a finite system with an edge?
Naively, it appears that the Floquet operator $U(T) = 1$ describes the trivial stroboscopic dynamics of a system with an effective Hamiltonian $H_{\rm eff} = 0$.
However, using the strip geometry shown in Fig.\ref{figSquare}b it is straightforward to check that the system supports chiral propagating edge modes localized on the boundaries.
Over each complete driving cycle, a particle moves by one unit cell along the edge.
These modes appear in the spectrum as two linearly-dispersing branches with group velocities $d\varepsilon/dk = \pm 1/T$, see Fig.\ref{figSquare}c.

Now consider the time-reversed cycle ($5-4-3-\cdots$), which has the opposite chirality.
Here, although particles circle around the plaquettes in the anti-clockwise direction, the Floquet operator (which is still identity) is the same as that of the original cycle.
Thus the information about the circulation direction, which is contained in the bulk evolution operator for intermediate times within the driving cycle, is absent in the Floquet operator.
Physically, this information is of crucial importance, however, as it determines the propagation direction of the edge states when the system is terminated.
Thus it is clear that a full topological characterization of these new phases requires an invariant defined in terms of the evolution operator evaluated throughout the entire driving period.

\section{Construction of the invariant}\label{sec:Analysis}
In this section we define an integer invariant which can be used to correctly predict the edge state
spectrum for a two dimensional periodically driven system. We begin by setting up the problem.
Let us consider a periodically driven, tight binding model defined on a 2D lattice with $N$ sites per unit
cell -- i.e.~a generalization of (\ref{Hexample}). Suppose that the hopping amplitudes are translationally
invariant and have finite range. Working in $\vec{k}$-space, the Hamiltonian can be written as
\begin{equation}
H(t) = \sum_{\vec{k} \alpha \alpha'} c^{\dagger}_{\vec{k}, \alpha} H_{\alpha\alpha'}(\vec{k}, t) c_{\vec{k}, \alpha'},
\label{Hperiod}
\end{equation}
where $\alpha,\alpha' = 1, ..., N$ label the sites in each unit cell and $\vec{k}$ lies in the first Brillouin zone.
All of the bulk properties of the system are encoded in the $N \times N$ Hermitian matrix $H(\vec{k},t)$.
In particular, the bulk time evolution operator can be computed as
\begin{equation}
U(\vec{k},t) = \mathcal{T} \exp \left(-i\int_0^t dt' \ H(\vec{k},t') \right),
\label{bulktimeevol}
\end{equation}
while the bulk Floquet operator corresponds to the special case $U(\vec{k},T)$.

Let us suppose that the Floquet spectrum has a gap extending over some finite interval
$[\varepsilon -\Delta \varepsilon, \varepsilon + \Delta \varepsilon]$.
Then, if we define the model in a geometry with an edge, any Floquet eigenstates with eigenvalues lying in this interval
must be localized near the boundary. These eigenstates correspond to edge modes. In analogy with the time-independent
case, we expect that the number of these edge modes -- counted with a sign corresponding to their chirality -- is
completely determined by the \emph{bulk} time evolution operator $U$. Therefore, one should be able to compute
the number of chiral edge modes at quasi-energy $\varepsilon$ given only $U(\vec{k},t)$.
We will now construct an explicit formula, defined in terms of $\{U(\vec{k},t), \varepsilon\}$, which gives exactly this
number. (Here $\{U(\vec{k},t)\}$ denotes the {\it set} of evolution operators for all times within a driving period,
$0 \le t \le T$).

\subsection{The case of a trivial Floquet operator}
Our construction proceeds in two steps. First, we consider the special case where the (bulk)
Floquet operator is simply the identity, i.e.~$U(\vec{k},T) = \vec{1}$ for all $\vec{k}$, as in the example described in Sec.~\ref{sec:Square}.
We then generalize to arbitrary Floquet operators.

If the Floquet operator is the identity, then
the Floquet spectrum is gapped everywhere except at $\varepsilon = 0$. Therefore, to each $\{U(\vec{k},t)\}$ we should be able to unambiguously
associate an integer $n_{\text{edge}}$ which counts the number of edge modes
propagating across the gap (i.e.~winding around the quasi-energy Brillouin zone). Furthermore,
this integer must be invariant under smooth deformations of $U$ that preserve the condition
$U(\vec{k},T) = \vec{1}$, since the number of edge modes cannot change under a deformation
in which the gap remains open.

Purely mathematical considerations suggest a natural guess: notice that $U$ is periodic in $k_x, k_y$
and $t$ (since $U(0) = U(T) = \vec{1}$ by assumption). Thus, $U$ defines a map from
$S^1 \times S^1 \times S^1 \rightarrow U(N)$.
Such maps are known to be classified by an integer topological
invariant or ``winding number'' defined by\cite{Bott78}
\begin{align}
W[U] = \frac{1}{8\pi^2} &\cdot \int dt dk_x dk_y \nonumber \\
&\cdot {\rm Tr} \left( U^{-1} \partial_{t} U
\cdot [U^{-1} \partial_{k_x} U , U^{-1} \partial_{k_y} U ] \right). \label{N[U]}
\end{align}
It is natural to guess that $n_{\text{edge}}$ is related to $W[U]$ in some way. In appendix \ref{edgewindapp}
we show that this guess is correct -- in fact, the two integers are identical:
\begin{equation}
n_{\text{edge}} = W[U].
\label{edgewindform}
\end{equation}

The winding number $W[U]$, defined in Eq.~(\ref{N[U]}), differs crucially from the familiar Chern number invariant. 
The winding number depends on the full {\it time evolution}, throughout the driving cycle, through the unitary evolution operator $U(t)$.
In contrast, the Chern number depends only on 
{\it projectors} onto a band of Floquet states 
(see Eq.~(\ref{chern2}) below).
The relationship between the winding numbers and the Chern numbers of the Floquet bands is discussed in more detail in Sec.~\ref{sec: relation}.

\subsection{The general case}
Next we consider the general case, where the Floquet operator $U(\vec{k},T)$ can be arbitrary.
We would like to compute the number of edge modes at a quasi-energy value $\varepsilon$ lying within a Floquet gap.
One way to do this is to reduce the problem to the previous case.
The idea is to construct another time evolution operator $U_\varepsilon$ satisfying
several properties. The first property is that $U_\varepsilon$ has a trivial Floquet operator:
$U_\varepsilon(\vec{k},T) = \vec{1}$
for all $\vec{k}$. The second property is that there exists a one-parameter family of evolution operators
$\{U_s : s \in [0,1]\}$ that smoothly interpolates between $U$ and $U_\varepsilon$:
\begin{align}
U_{s=0}(\vec{k},t) = U(\vec{k},t) \ , \ \ U_{s=1}(\vec{k},t) = U_\varepsilon(\vec{k},t).
\end{align}
Finally, and most importantly, we require that the interpolation $U_s(\vec{k},T)$
maintains a gap around some quasi-energy value $\varepsilon_s$ which changes smoothly in $s$ and satisfies
$\varepsilon_{s=0} = \varepsilon$ and $\varepsilon_{s=1} = \pi/T$.

If we can construct an evolution operator $U_\varepsilon$ with these properties then we can
immediately compute the number of chiral edge modes of $U$ at quasi-energy $\varepsilon$:
\begin{equation}
n_{\text{edge}}(\varepsilon) = W[U_\varepsilon].
\label{genwindform}
\end{equation}
The validity of this identification comes from the fact that the gap doesn't close during the interpolation
process (by assumption).
Therefore, the number of edge modes of $U$ at quasi-energy $\varepsilon$ must be the same as the number of
edge modes of $U_\varepsilon$ at quasi-energy $\pi/T$. The latter quantity is then given by $W[U_\varepsilon]$,
using the formula in Eq.~(\ref{edgewindform}).

All that remains is to construct an appropriate $U_\varepsilon$ and a corresponding interpolation.
There is some arbitrariness here, since $U_\varepsilon$ is far from unique. The result, however,
will not depend on our particular choice. We will define $U_\varepsilon$ by
\begin{equation}
U_\varepsilon(\vec{k},t) =
\begin{cases} U(\vec{k}, 2t)
& \mbox{if } 0 \leq t \leq T/2 \\
V_\varepsilon(\vec{k},2T-2t) & \mbox{if } T/2 \leq t \leq T,
\end{cases}
\end{equation}
where
\begin{align}
V_\varepsilon(\vec{k},t) = e^{-i H_{\rm eff}(\vec{k}) t } \ , \ H_{\rm eff}(\vec{k}) = \frac{i}{T} \log U(\vec{k},T).
\label{Vu}
\end{align}
Here, we choose the branch cut of the logarithm to lie along the direction $e^{-i\varepsilon T}$. That is,
we choose a branch with
\begin{eqnarray}
\log e^{-i \varepsilon T + i 0^-} &=& -i \cdot \varepsilon T \nonumber \\
\log e^{-i \varepsilon T + i 0^+} &=& -i \cdot \varepsilon T -2\pi i.
\label{branch}
\end{eqnarray}
This choice of branch is important, and provides the only dependence of $U_\varepsilon$ on $\varepsilon$.

Physically, $V_\epsilon$ can be viewed as a trivial ``return map'' which is used
to connect the Floquet operator $U(\vec{k},T)$ to the identity.
The quantity $H_{\rm eff}$ appearing in the exponent plays the role of a static effective Hamiltonian which generates a bulk time evolution that, when examined stroboscopically at integer multiples of the driving period $T$, is identical to that of
$U(\vec{k},T)$\cite{KitagawaPRB}.
By concatenating the driving cycle with an evolution generated by $-H_{\rm eff}$, the net effect of
the combined evolution becomes trivial.

We now check that the above definition of $U_\varepsilon$
satisfies all of our requirements. It is clear
that $U_\varepsilon(\vec{k},T) = \vec{1}$; the crucial question is to find an appropriate interpolation connecting
$U$ and $U_\varepsilon$. The following interpolation does the job:
\begin{equation}
U_s(\vec{k},t) =
\begin{cases} U(\vec{k}, (1+s)t)
& \mbox{if } 0 \leq t \leq T/(1+s) \\
V_{\varepsilon}(\vec{k},2T-(1+s)t) & \mbox{if } T/(1+s) \leq t \leq T.
\end{cases} \nonumber
\end{equation}
Indeed, using Eq.~(\ref{Vu}) it is easy to check that the Floquet gap remains open around
$\varepsilon_s \equiv (1-s)(\varepsilon+\pi/T) - \pi/T $, using the fact that
\begin{equation}
U_s(\vec{k},T) = U(\vec{k},T)^{1-s}.
\end{equation}
Also, we can see that $\varepsilon_{s=0} = \varepsilon$ and $\varepsilon_{s=1} = \pi/T$ (mod $2\pi/T$). Thus, the interpolation
$U_s$ satisfies all of the conditions listed above.

\subsection{Relation with Chern number}
\label{sec: relation}
As discussed in Sec.~\ref{sec:Failure}, the Chern number of a Floquet band $n$ is defined by integrating the Berry curvature of the Floquet eigenstates $|\psi_{n}(\vec{k}, t)\>$ over the crystal momentum Brillouin zone, at a fixed time $t$:
\begin{equation}
\mathcal{C}_n = -\frac{1}{2\pi} \int dk_x dk_y \ (\vec{\nabla} \times \vec{\mathcal{A}}_n),
\label{chern1}
\end{equation}
where $\vec{\mathcal{A}}_n = \<\psi_{n}(\vec{k}, t)| i \vec{\nabla} |\psi_{n}(\vec{k}, t)\>$. Equivalently, $\mathcal{C}$
can be written as
\begin{equation}
\mathcal{C}_n = \frac{1}{2\pi i} \int dk_x dk_y \ {\rm Tr}\left(P_n \cdot [\partial_{k_x} P_n, \partial_{k_y} P_n] \right),
\label{chern2}
\end{equation}
where $P_n(\vec{k}) = |\psi_{n}(\vec{k}, t)\>\<\psi_{n}(\vec{k}, t)|$ is a projector onto the Floquet eigenstate $|\psi_{n}(\vec{k}, t)\>$.

Not surprisingly, there is a close mathematical relationship between the winding number $W[U_\varepsilon]$ defined above
and the Chern numbers $\{\mathcal{C}_n\}$ of the Floquet bands. This relationship becomes clear when one considers the
\emph{difference} between winding numbers evaluated at two different quasi-energies. More specifically,
it is possible to show that
\begin{equation}
W[U_{\varepsilon'}] - W[U_\varepsilon] =  \mathcal{C}_{\varepsilon \varepsilon'},
\label{windchernrel}
\end{equation}
where $\mathcal{C}_{\varepsilon \varepsilon'}$ denotes the 
sum of the Chern numbers of all Floquet bands that lie in between
$\varepsilon$ and $\varepsilon'$ (see appendix \ref{edgechernapp} for a derivation). This identity is very natural from
a physical point of view. Indeed, identifying $W[U_\varepsilon]$ with
$n_{\text{edge}}(\varepsilon)$, Eq.~(\ref{windchernrel}) is simply the statement that difference between the numbers of
edge modes at two quasi-energies $\varepsilon$ and $\varepsilon'$ is equal to the total Chern number of the intermediate bands.

\subsection{Phases of the modulated square lattice model}
\begin{figure}[t]
\includegraphics[width=0.95\columnwidth]{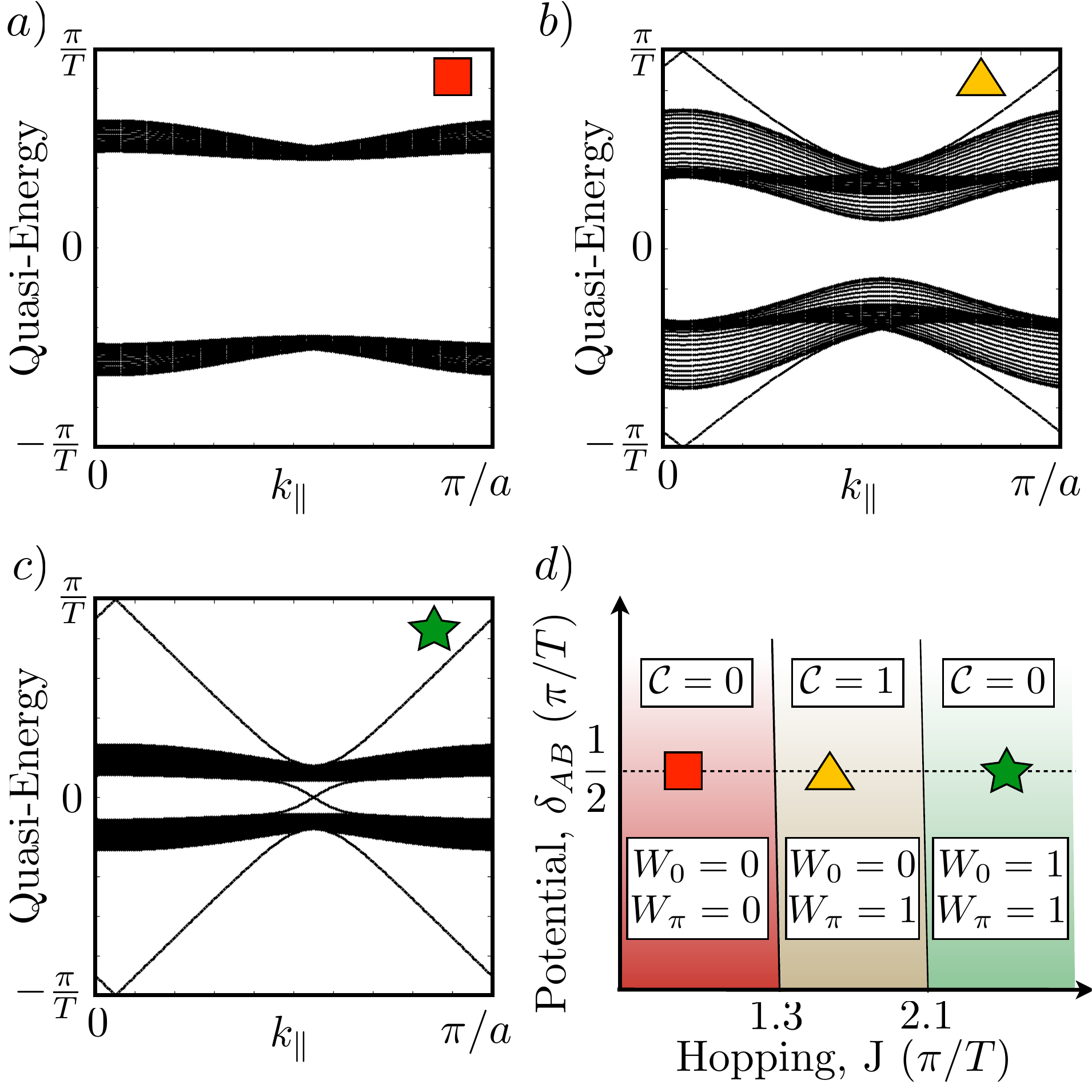}
 \caption[]{Phases of the square lattice model in Sec.~\ref{sec:Square}.
a-c) Example Floquet spectra for each of the three phases, calculated in a strip geometry (unit cell $2a$), with
sublattice potential $\delta_{AB} = 0.5\, \pi/T$ and hopping amplitudes a) $J = 0.5\, \pi/T$, b) $J = 1.5\, \pi/T$
and c) $J = 2.5\, \pi/T$. The winding numbers $W_0$ and $W_\pi$ correctly yield the numbers of edge
states in the two gaps. The asymmetry in $k_\parallel$ is due to the breaking of inversion symmetry by the sublattice potential $\delta_{AB}$.
d) Phase diagram indicating the winding numbers $W_0$ and $W_\pi$,
calculated in the gaps at quasi-energies 0 and $\pi/T$, and Chern number $\mathcal{C}$ of the upper band.
Note the existence of two topologically-distinct phases with $\mathcal{C} = 0$.
}
\label{figPhaseDiagram}
\end{figure}

We now briefly illustrate the utility of the winding number by examining the phase diagram of the model introduced in Sec.~\ref{sec:Square}. 
Away from the special point $\delta_{AB} = 0$, $JT/5 = \pi/2$, the Floquet operator $U(T)$ is not equal to the identity.
Thus the return-map construction described above must be used to compute the winding numbers for the Floquet gaps around quasi-energy values 0 and $\pi$.

In the general case, the evolution operator $U(t)$ for $0 \le t \le T$ is straightforward to calculate.
The Hamiltonian in Eq.~(\ref{Hexample}) is piecewise constant in time through five equal-length segments of duration $T/5$.
Let $H_n$ denote the Hamiltonian within the interval $(n-1)T/5 \le t \le nT/5$. 
Then for $t < T/5$, we have $U(t) = \exp({-iH_1t})$;
for $T/5 \le t < 2T/5$, $U(t) = e^{-iH_2(t-T/5)}e^{-iH_1T/5}$, and so on.
Because the Bloch Hamiltonians $\{H_n(\vec{k})\}$ are $2\times 2$ matrices, their exponentials are easily obtained directly. 
The evolution operator $U(\vec{k},t)$ is then given by a handful of $2\times 2$ matrix exponentiations and multiplications.
The return map $V_\epsilon(\vec{k},t)$, Eq.~(\ref{Vu}), is defined through the logarithm of $U(\vec{k},T) = e^{-iH_5T/t}\cdots e^{-iH_1T/5}$.
Using these results, the winding number $W[U_\epsilon]$, see Eqs.~(\ref{N[U]}) and (\ref{genwindform}), can then be calculated explicitly.

Using direct numerical calculation, we solved for Floquet spectra in both periodic and strip geometries (see e.g.~Figs.~\ref{figPhaseDiagram}a-c), and mapped out the approximate phase diagram, shown in Fig.~\ref{figPhaseDiagram}d. 
Example spectra from each of the three phases confirm the correspondence between the winding number, calculated from the bulk evolution operator in the periodic geometry, and the edge state count in each gap in the strip geometry. 
Figure \ref{figPhaseDiagram}c provides an explicit example of the phenomenon discovered in Ref.~\onlinecite{KitagawaPRB} and shown in Fig.~\ref{fig1}: chiral edge modes appear for the finite system, despite the fact that the Chern numbers of both bands are zero.

\section{Frequency domain formulation}
\label{sec:Nate}
In this section we discuss an alternative approach for deriving
the edge state spectrum of the Floquet operator, and its relation
to the Chern numbers of the bulk Floquet bands. This approach
is based on an analysis in the \emph{frequency} domain instead of
the time domain.
The information obtained from the frequency domain analysis is equivalent to that obtained from the winding numbers 
in Sec.~\ref{sec:Analysis}.
Specifically, both methods predict the number of chiral edge modes in each Floquet gap.
In practice, the frequency domain technique provides a simpler computational route for systems in which the driving field is weak and has a narrow power spectrum. 

\subsection{Repeated zone analysis}
We start from the Schr\"{o}dinger equation with Hamiltonian
(\ref{Hperiod}) for the Floquet state in band $n$, with crystal
momentum $\vec{k}$. Using a basis of states labeled by $\alpha$,
we write $\Ket{\psi_n(\vec{k},t)} =
\sum_{\alpha=1}^{N}
\psi_{n\alpha}(\vec{k},t)c^\dagger_{\vec{k},\alpha}\Ket{0}$, where
$\Ket{0}$ is the vacuum. The amplitudes
$\{\psi_{n\alpha}(\vec{k},t)\}$ evolve according to:
\begin{equation}
  i\partial_t\psi_{n\alpha}(\vec{k},t) = \sum_{\alpha'=1}^{N}H_{\alpha\alpha'}(\vec{k},t)\psi_{n\alpha'}(\vec{k},t).
\end{equation}

Employing the Floquet theorem, we write
\begin{equation}
\psi_{n\alpha}(\vec{k},t) = e^{-i\epsilon_n(\vec{k})t}\!\sum_{m = -\infty}^\infty \varphi_{n\alpha}^{(m)}(\vec{k})e^{im\omega t},
\label{eq:time_dep_sol}
\end{equation}
 where $\omega = 2\pi/T$. Below we suppress all $\vec{k}$ indices for notational simplicity.
The coefficients $\varphi_{n\alpha}^{(m)}$ satisfy the (time-independent) eigenvalue equation 
\begin{equation}
\label{FloquetSE}\sum_{\alpha',m'}\mathcal{H}_{\alpha\alpha'}^{mm'}\varphi_{n\alpha'}^{(m')} = \epsilon_n \varphi_{n\alpha}^{(m)},
\end{equation}
where the ``Floquet Hamiltonian'' $\mathcal{H}_{\alpha\alpha'}^{mm'}$ is given by
\begin{equation}
\label{FloquetH}
\mathcal{H}_{\alpha\alpha'}^{mm'} = m \omega \delta_{\alpha \alpha'}\delta_{mm'} + \frac{1}{T}\int_0^T dt\,e^{-i(m-m')\omega t} H_{\alpha\alpha'}(t).
\end{equation}

For each $\vec{k}$, Eq.~(\ref{FloquetSE}) has solutions throughout
$-\infty < \epsilon_n < \infty$. However, as discussed in
Sec.~\ref{sec:Failure}, if $\epsilon_n$ is an eigenvalue of
Eq.~(\ref{FloquetSE}) corresponding to the eigenstate with amplitudes
$\{\varphi_{n\alpha}^{(m)}\}$, then $\tilde{\epsilon}_n = \epsilon_n + p\omega$ (where $p$ is any integer) is also an eigenvalue corresponding to an eigenstate with amplitudes given by 
$\tilde{\phi}_{n\alpha}^{(m+p)} =
\phi_{n\alpha}^{(m)}$.
In fact, as seen by direct substitution into Eq.~(\ref{eq:time_dep_sol}), 
all of these solutions correspond to the {\it same} time-dependent
solution of the Schrodinger equation.
Therefore, the Floquet states are uniquely and completely parametrized by quasi-energies in the ``first quasi-energy Brillouin zone,'' $-\pi/T < \epsilon_n < \pi/T$. 
Equation (\ref{FloquetSE}) is the temporal analogue of a ``repeated zone'' scheme for conventional band structure calculations.


\begin{figure}[t]
\includegraphics[width=0.8\columnwidth]{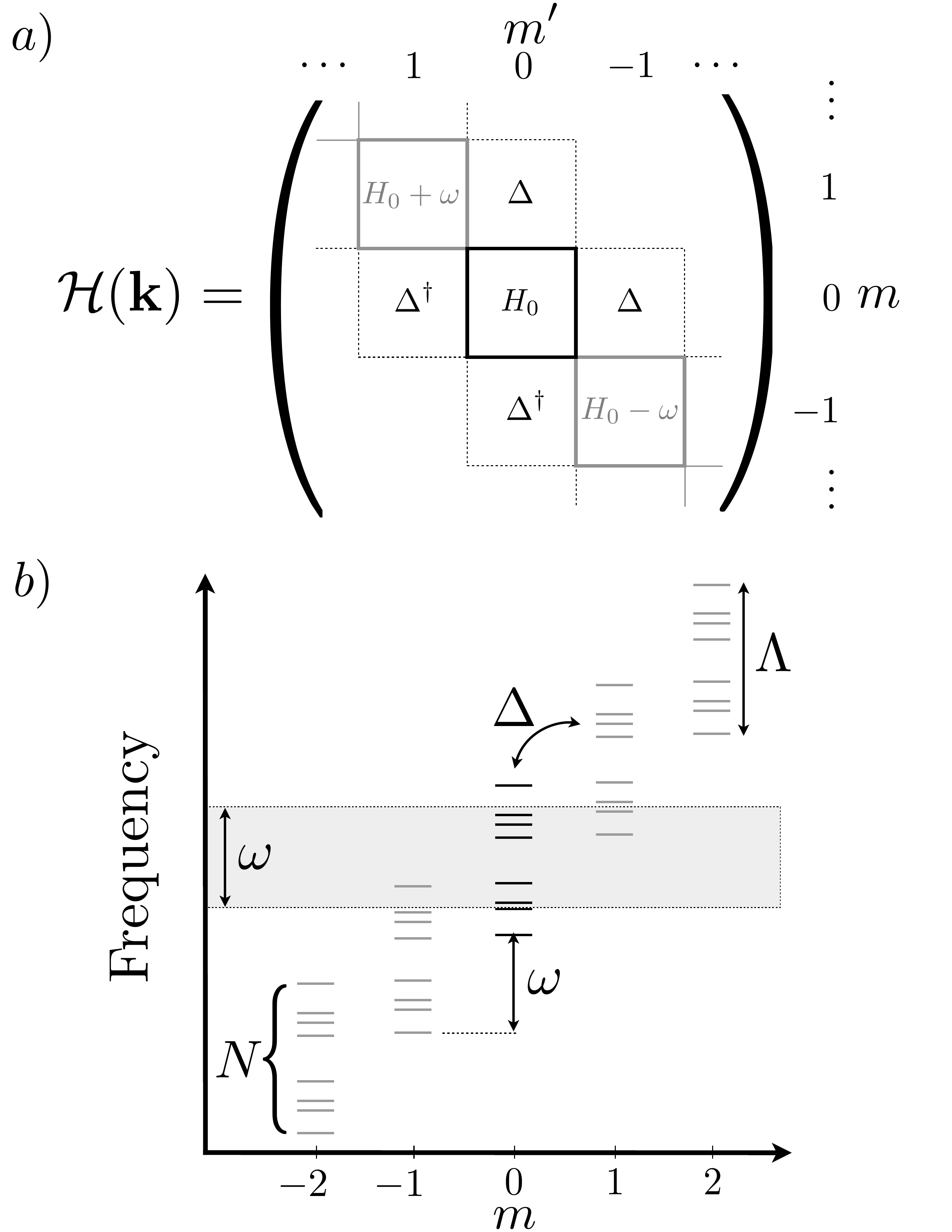}
 \caption[]{Floquet Hamiltonian and level diagram in the repeated zone scheme.
a) When the driving $\Delta(t)$ consists of a single harmonic with angular frequency $\omega$, the Floquet Hamiltonian $\mathcal{H}(\vec{k})$, see Eq.~(\ref{FloquetH}), assumes a block-tridiagonal form.
Each block is an $N\times N$ matrix, where $N$ is the number of bands in the unperturbed Hamiltonian $H_0$.
The off-diagonal blocks labeled by $\Delta$ and $\Delta^\dagger$ describe transitions accompanied by the absorption and emission of a driving field photon, respectively.
b) Schematic Floquet level diagram.
The $N$-level spectrum of $H_0(\vec{k})$, which spans an energy range $\Lambda$, is copied and rigidly shifted by $m\omega$ for each value of $m$.
The harmonic driving $\Delta$ induces transitions between neighboring copies with $m$-values differing by 1.
Analogous to Wannier-Stark states, the Floquet eigenstates in the first quasi-energy Brillouin zone (gray shaded region) are localized in $m$, with amplitudes highly suppressed for $|m| \gg \Lambda/\omega$.
}
\label{figRepeatedZoneMatrix}
\end{figure}

To illustrate the structure of Eqs.~(\ref{FloquetSE}) and (\ref{FloquetH}),
we now consider the case where
\begin{equation}
H(t) = H_0 + \Delta e^{i\omega t} + \Delta^{\dagger} e^{-i \omega t}.
\label{harmdriving}
\end{equation}
In this case, the matrix $\mathcal{H}^{mm'}_{\alpha\alpha'}$ has the block tri-diagonal form shown schematically in Fig.~\ref{figRepeatedZoneMatrix}a, where each block is an $N\times N$ matrix. As noted above, it suffices to solve for the
Floquet bands within the first quasi-energy Brillouin zone, $-\pi/T < \epsilon_n < \pi/T$.
In the limit of weak driving (small $\Delta$), this can be done in perturbation theory.


As sketched in Fig.~\ref{figRepeatedZoneMatrix}b, the zeroth-order spectrum of the Floquet
Hamiltonian (\ref{FloquetH}) consists of an array of copies of the original spectrum of $H_0$,
shifted up and down by integer multiples of the drive frequency $\omega$.
In the Floquet picture, the harmonic driving $(\Delta e^{i\omega t} + \text{h.c.})$ induces hopping between levels with $m$-values differing by one.
Processes where $m$ increases (decreases) by one correspond to transitions accompanied by the absorption (emission) of a photon from the driving field.
When the range $\Lambda$ of the spectrum of $H_0$ is greater than $\omega$, resonant transitions
(associated with degeneracies between levels with neighboring values of $m$) may occur for 
some values of $\vec{k}$.
To leading order, degenerate perturbation theory shows that the driving $(\Delta e^{i\omega t} + \text{h.c.})$ opens gaps at these resonances (see Fig.\ref{figRepeatedZoneBands}b).

\subsection{Truncation of the Floquet Hamiltonian and edge states}
An important property of Floquet eigenstates is that they are {\it localized} in $m$, decaying rapidly
for $|m-m_0|\omega \gg \Lambda$, where $m_0$ is the center of a given localized state. This localization
in frequency space is analogous to the well-known localization of Wannier-Stark states in real space.
Starting from this observation, we now describe an alternative method for computing edge state spectra of
Floquet systems -- complementary to the winding number approach of Sec.~\ref{sec:Analysis}.

The first step is to truncate the Floquet Hamiltonian
(\ref{FloquetH}) so that $m$ and $m'$ run over a large but finite
range, $-M \leq m,m' \leq M$, where $M$ is much greater than the
frequency-space localization range of the Floquet states. Making
use of the localization of the Floquet states in $m$, we note that
the spectrum of the truncated Floquet Hamiltonian will be a good
approximation to the exact result within the first few
quasi-energy zones centered around $m = 0$. In particular,
if we consider a geometry with a boundary, then the
{\it edge state spectrum} of the truncated Hamiltonian will closely
match the exact Floquet edge state spectrum within these central few quasi-energy zones.

The key observation is that there is actually a straightforward
way to determine the edge state spectrum using Chern numbers of
the bands of the {\it truncated} Floquet Hamiltonian. Interpreting
the truncated Floquet Hamiltonian as the static Hamiltonian of
some new $(2M + 1)\times N$-band system, we apply the standard
bulk-edge correspondence to deduce that the net number of
chiral edge modes (counterclockwise minus clockwise) crossing any
particular gap is given by the sum of the Chern numbers of all
bands below this gap. In this way, we have a simple procedure for
deriving the edge state spectra of Floquet systems: to find the
number of chiral edge modes crossing a given quasi-energy gap
within the first quasi-energy Brillouin zone, i.e.~within the interval
$-\pi/T < \varepsilon < \pi/T$, we truncate the Floquet
Hamiltonian and then count Chern numbers of the resulting bands
below the gap in which we are interested. Note that, as in Sec.~\ref{sec: relation}, the difference between the numbers of
chiral edge modes at quasi-energies $\varepsilon$ and
$\varepsilon'$ within the first quasi-energy zone is equal to the sum of the
Chern numbers of all Floquet bands between $\varepsilon$ and
$\varepsilon'$.

\begin{figure}[t]
\includegraphics[width=\columnwidth]{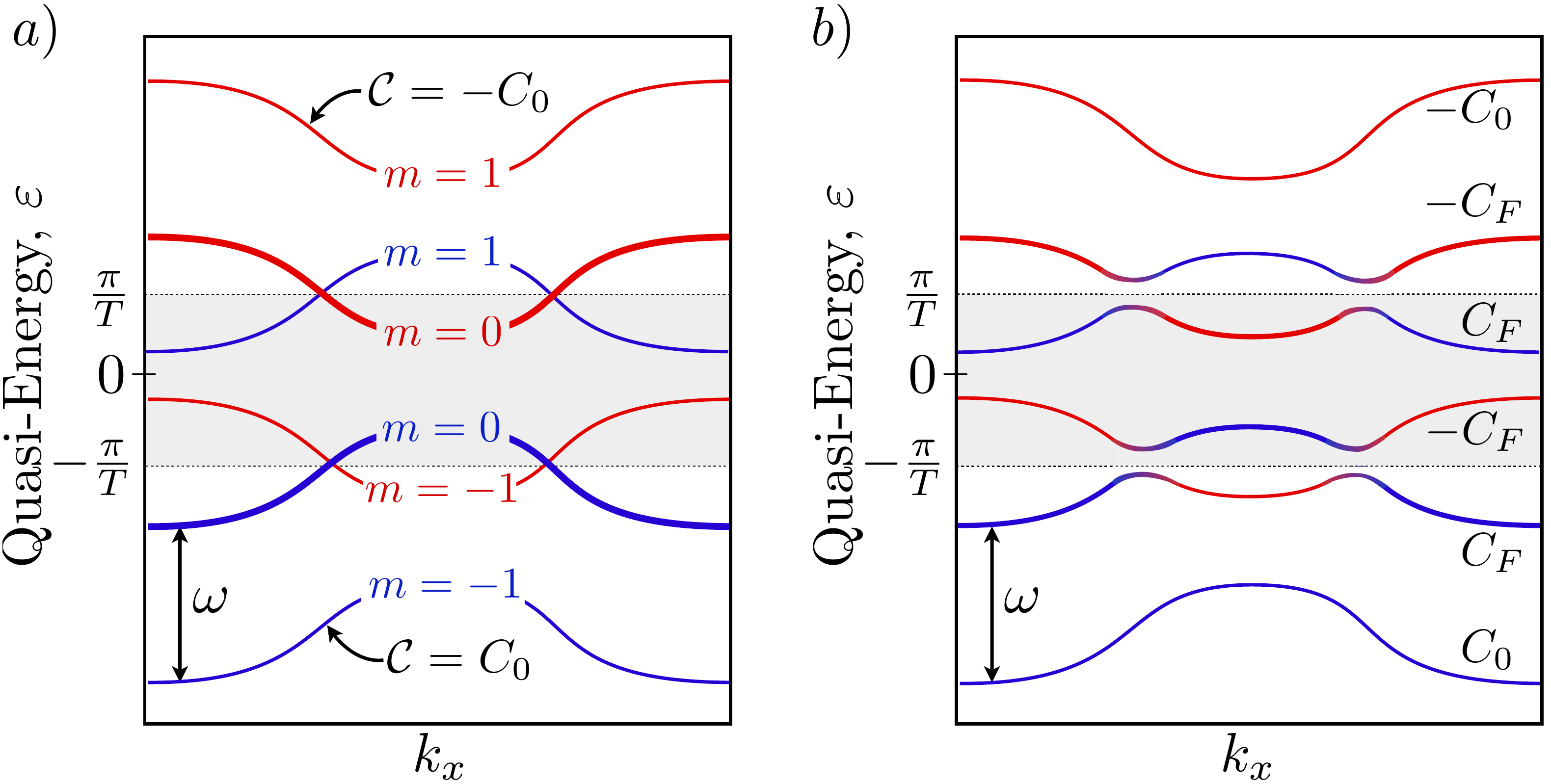}
 \caption[]{Floquet bands for a two-band model in the truncated repeated zone scheme.
Representative one-dimensional cuts through the two-dimensional crystal momentum Brillouin zone are shown.
a) For $\Delta = 0$, the unperturbed bands of $H_0$ are each copied and shifted up and down one time, $m = -1, 0, 1$.
The bands have Chern numbers $\mathcal{C} = \pm C_0$. 
The crossings between bands with $m$-values differing by 1 correspond to points in the Brillouin zone where the driving field resonantly couples the two bands.
b) For $\Delta \neq 0$, the driving field opens avoided crossings at the resonance points.
After coupling, the Chern numbers of the crossed bands are changed to $\mathcal{C} = \pm C_F$.
In the truncated scheme,
the 
appearance of Floquet edge states at a particular quasi-energy in the first quasi-energy Brillouin zone (gray region) can be predicted by adding up the Chern numbers of all bands below it.
}
\label{figRepeatedZoneBands}
\end{figure}

As an example, we study the case of harmonic driving introduced in Eq.~(\ref{harmdriving}).
We take $H_0$ to be a general two-band Hamiltonian,
\begin{equation}
H_0(\bk)=e(\bk)+\bd(\bk)\cdot\boldsymbol{\sigma},
\label{eq:twoband}
\end{equation}
where $\boldsymbol{\sigma} = (\sigma_x,\sigma_y,\sigma_z)$ is the vector of Pauli matrices, and
$\bd(\bk)$ is a three dimensional vector. The bands are
characterized by Chern numbers $\mathcal{C} =\pm C_0$, as indicated in Fig.~\ref{figRepeatedZoneBands}a. For
simplicity we assume that the periodic drive creates a single
resonance between the valence and conduction bands of $H_0$. This
resonance occurs simultaneously for all $\bk$-values which satisfy
$\omega = 2|\bd(\bk)|$. 
We further assume that this resonance condition is satisfied along
a single closed curve in the crystal momentum Brillouin zone (see
Fig.~\ref{fig:numerical_trunc}a).

Consider the corresponding Floquet Hamiltonian
$\mathcal{H}_{\alpha\alpha'}^{mm'}$, see Eq.~(\ref{FloquetH}),
which is truncated to the range $-1 \le m,m' \le 1$. The band
structure of the truncated Floquet Hamiltonian is shown
schematically in Fig.~\ref{figRepeatedZoneBands}.
For $\Delta=0$, the valence band of block $m$ and the conduction band from block $m-1$ 
are degenerate on the resonance curve (appearing as two points on the sections shown in Fig.~\ref{figRepeatedZoneBands}).
The driving terms mix the two bands, opening gaps at the crossing points. 
The resulting bands have Chern numbers $C_F$ and $-C_F$, as
indicated in Fig.~\ref{figRepeatedZoneBands}b.  Due to the
truncation, however, two isolated bands are left, at the top and
bottom of the spectrum. These bands do not participate in any
resonances, and their Chern numbers remain equal to $\pm C_0$, see
Fig.~\ref{figRepeatedZoneBands}b.

Examining the gap centered at quasi-energy $-\pi/T$, we see that
the Chern numbers of all bands below it sum to $C_0+C_F$.
Therefore we expect to find $C_0 + C_F$ chiral edge modes spanning
this gap. For the same reason, the gaps centered at quasi-energy
values $\epsilon = 0$ and $\epsilon =\pi/T$  support $C_0$ and
$C_0+C_F$ chiral edge modes, respectively.

The value of $C_F$ which is obtained in any particular model depends on the details of 
$H_0(\bk)$, $\Delta$, and $\omega$. In particular, in
Sec.~\ref{sec: zero chern one res} below and in
Fig~\ref{fig:numerical_trunc} we describe a specific model which gives $C_F=0$, while nonetheless exhibiting robust chiral edge modes.

Importantly, the method outlined above for determining the number
of chiral edge modes in each gap of the Floquet spectrum can be
applied to more general periodically driven systems, and is not
restricted to the weakly driven, single harmonic limit discussed
in the above example. It works provided that one keeps
sufficiently many copies of $H_0$ in the truncation, and that the
driving power spectrum decays sufficiently rapidly for higher
harmonics. If the driving is strong, the relationship between the
Floquet bands and the original bands may be more complicated than
in the example considered above, but the results of this procedure
are still guaranteed to converge for large enough $M$.

\subsection{Anomalous edge states in a weakly-driven two-band lattice system} 
\label{sec: zero chern one res}
\begin{figure}[t]
\includegraphics[width=1\columnwidth]{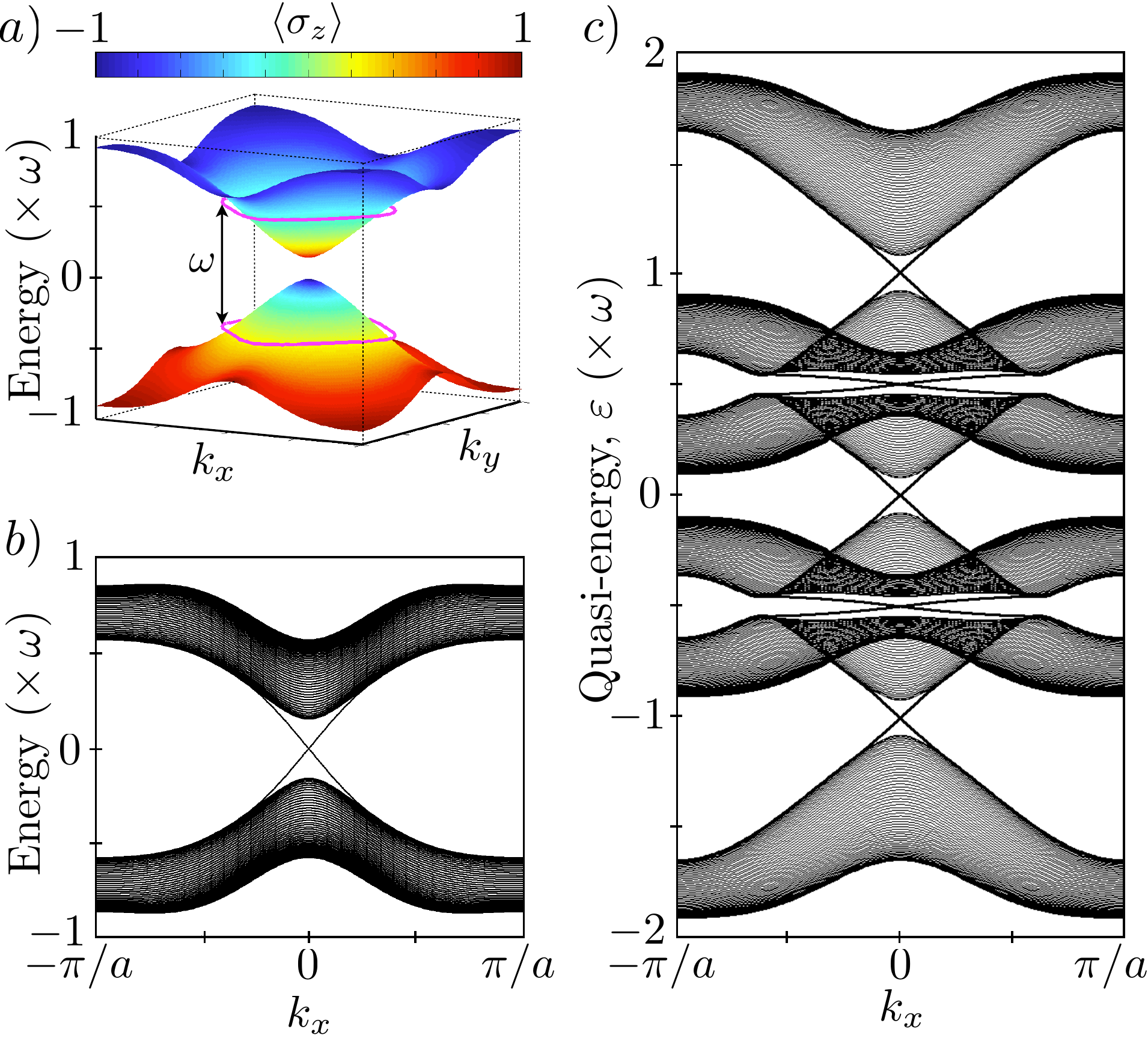}
 \caption[]{Anomalous edge states in a weakly-driven two-band system, with Hamiltonian $H_0$ given by Eqs.~(\ref{eq:twoband}) and (\ref{eq:model2param}).  The Chern numbers of the unperturbed bands are $C_0=\pm1$.
 a) Spectrum of $H_0$ with periodic boundary conditions, showing the resonance curve (purple line) in the valence and conduction
 bands. The color scheme indicates $\Avg{\sigma_z}$ for the corresponding eigenstates in the two bands. 
 b) Spectrum of $H_0$ 
for a cylindrical geometry, with periodic boundary conditions in the $x$-direction and open boundary conditions in the $y$-direction.
   c) Spectrum of the truncated Floquet Hamiltonian, Eq.~(\ref{FloquetH}), in the cylindrical geometry.
  The isolated bands at the top and bottom of the figure 
  do not participate in any resonances and are nearly identical to 
  the original bands of $H_0$.
  The Floquet bands have zero Chern numbers, while edge states traverse both gaps, 
 centered around $\varepsilon=0$ and $\varepsilon=\pi/T$.
The parameters used are $J/\mu=b/\mu=1.5$, $a/\mu=4$,
$\Delta_0/\mu=1$, and $\Delta_0/\omega=0.07$.}
\label{fig:numerical_trunc}
\end{figure}

We consider a two-band model with a Hamiltonian of the
form~(\ref{eq:twoband}), with $e(\bk) = 0$ and with the components
of $\bd(\bk)$ given by:
\begin{eqnarray}
\nonumber d_x(\bk) &=& a \sin(k_x),\quad d_y(\bk) = a\sin(k_y)\\
d_z(\bk)&=&(\mu-J)-2b(2-\cos(k_x)-\cos(k_y))\nonumber\\
 &\phantom{=}&+ J \cos(k_x)\cos(k_y).
 \label{eq:model2param}
\end{eqnarray}
In the parameter regime $\mu,b,J>0$ and $\mu/b<4$, the unperturbed
bands of $H_0$ have Chern numbers $C_0=\pm 1$, as indicated in
Fig.~\ref{fig:numerical_trunc}a. This result can easily be
obtained by direct integration of the Berry curvature using
Eq.~(\ref{chern1}). More simply, recall that for a general
two-band system of the form (\ref{eq:twoband}), the Chern number
is determined by the degree of the map
$\hbd(\bk)=\bd(\bk)/|\bd(\bk)|$ from the Brillouin zone to the
unit (Bloch) sphere $S^2$, i.e.~by the number of times this map
``covers'' $S^2$. The result $C_0=\pm 1$ follows simply by noting
that the vector $\hbd(\bk)$ maps {\it exactly one point} in the Brillouin
zone to a point close to the north pole of the unit sphere, as
indicated by the colored shading in Fig.~\ref{fig:numerical_trunc}a (see
 Ref.~\onlinecite{Wang2010} for more details).

We now add a time-dependent perturbation of the form
$\Delta(t)=\Delta_0\sigma_z\cos(\omega t)$. The hopping $J$ and frequency
$\omega$ can be tuned such that only a single resonance occurs in
the Brillouin zone, on a closed curve around $\bk=0$ as shown in
Fig.~\ref{fig:numerical_trunc}a. Note that here we take parameters to provide the simplest situation, containing only a single resonance, for the purposes of clearest illustration; similar
considerations apply in other parameter regimes allowing for more
resonances.

To see that this driving results in ``trivial''
Floquet bands with $C_F = 0$, we perform degenerate perturbation
theory in the off-diagonal matrix elements of the Floquet
Hamiltonian, Eq.~(\ref{FloquetSE}).
Consider an isolated pair of crossing bands, such as the red and
blue bands labeled $m = -1$ and $m = 0$ in
Fig.~\ref{figRepeatedZoneBands}a, respectively. To lowest order in
$\Delta_0$, the corresponding two new hybridized bands are described as
the eigenstates of an effective Hamiltonian
\begin{equation}
H^{(0)}_{\rm eff} =
\left(|\bd(\bk)|-\omega/2\right)\hbd(\bk)\cdot\boldsymbol{\sigma}+\tilde{\bf \Delta}(\bk)\cdot\boldsymbol{\sigma},
\label{eq:Heff0}
\end{equation}
where $\tilde{\bf \Delta}(\bk)=\Delta_0[\hbz- \hat{d}_z(\bk)\hat{\bd}(\bk)]$.
Note that $H^{(0)}_{\rm eff}$ is in fact a leading-order
approximation to the effective Hamiltonian introduced in
Eq.~(\ref{Vu}).
As indicated in Fig.~\ref{figRepeatedZoneBands}b, the bands of $H_{\rm eff}^{(0)}$ are {\it inverted} relative to those of $H_0$ in the region around $\bk = 0$. 

How do we see that these bands have trivial Chern indices?
Consider the map from the Brillouin zone to $S^2$ defined by $H^{(0)}_{\rm
eff}$ in Eq.(\ref{eq:Heff0}). A necessary requirement for
obtaining a nonzero Chern number is that every point on the Bloch
sphere must be reached for at least one value of $\bk$ in the
Brillouin zone.
For small $\Delta_0$, and for $\omega$ such that the resonance curve winds around $\bk=0$, there is a region surrounding the north pole of $S^2$ which is not accessed by $H_{\rm eff}^{(0)}$ for {\it any} points in the Brillouin zone. 
Therefore its bands {\it must} have zero Chern numbers.

In Fig.~\ref{fig:numerical_trunc}c, we plot the spectrum of the truncated Floquet Hamiltonian for the model described above, defined in a cylindrical geometry. 
This figure clearly exhibits chiral edge states traversing each of the gaps, including those associated with the isolated bands at the top and bottom of the truncation window.
As described above, these bands are not strongly hybridized, and are nearly identical to the unperturbed bands of $H_0$ (see Fig.~\ref{fig:numerical_trunc}b). 

The analysis above provides an alternative picture for
understanding how edge states can appear in periodically-driven
systems, even when all of the Floquet bands have vanishing Chern
numbers. The fact that the bands carry zero Chern numbers only
guarantees that the numbers of chiral edge states in the two gaps
surrounding each Floquet band are equal. In the truncated Floquet
Hamiltonian picture, the edge states owe their existence to bands
near the truncation boundaries which carry anomalous
(i.e.~non-vanishing) Chern numbers. These bands retain the history
about the topological properties of the original bands of $H_0$,
as well as the structure of the resonances which transform them
into the Floquet bands. Note that anomalous edge states can be
obtained also in a model when the unperturbed Hamiltonian
$H_0(\bk)$ contains only {\it trivial} bands, provided that
additional resonances (e.g.~due to multi-photon processes) are allowed.\\



\section{Discussion}\label{sec:Concl}
In this paper, we discussed the correspondence between bulk
topological invariants and the edge state spectra of two dimensional
periodically driven systems. In both static and periodically-driven systems, the Chern number of a given band
is equal to the difference between the number of chiral edge
states above and below the band. Because the spectrum of a static
system is bounded from below, knowledge of the Chern numbers of
all bands up to a particular energy is sufficient to uniquely
determine the number of chiral edge modes at that energy. For
driven systems, however, the spectrum is defined on a circle, and
hence is {\it not} bounded. Knowing the Chern numbers of all
Floquet bands is therefore not sufficient to determine the edge
state structure. Our main result in this work 
is the construction of a new invariant which fully captures the
edge state spectrum for periodically driven systems.

The most striking example of the difference between the static and driven cases occurs when a driven system supports robust chiral edge states, even when the Chern numbers of all of its Floquet bands are zero. 
Can such systems be realized experimentally?
In the analysis above we introduced two lattice models: one with strongly modulated hopping amplitudes, and one with a weak driving field applied on top of a convenient band structure.
While the first model was constructed primarily for mathematical demonstration, a class of models similar to the second, involving weak, uniform driving, should be accessible in cold atomic, solid state, or all-photonic systems.


A promising route for implementing  models of the second type
is to utilize Cooper's optical flux-lattice systems\cite{Cooper2011}.
There, weak periodic modulations of the optical fields which create the flux lattice could be used to produce the resonances necessary for obtaining a non-trivial Floquet spectrum.
A modulated flux lattice with ``anomalous'' edge states could
be achieved either by starting with topologically-nontrivial unperturbed bands, as in Sec.~\ref{sec: zero chern one res}, or even
by starting with topologically-trivial bands and employing two (or more) photon resonances.

Solid state implementations are also possible (see e.g.~Refs.~\onlinecite{Vu2004,McIver12} for some experimental preliminaries). Typically, the
frequency $\omega$ will be at least a few times smaller than the total
width $\Lambda$ of the two central bands of interest (e.g.~the conduction
and valence bands). In this case, the Floquet band structure is
complicated by quasi-energy zone folding, which leads to
additional band crossings corresponding to multi-photon
resonances. Generically, for $\omega/\Lambda \ll 1$, one can find bulk states coming from a
sufficiently high-order folding at any quasi-energy. Therefore,
strictly speaking, there is no gap in the Floquet spectrum at any
quasi-energy. If there are chiral edge states in a gap which
corresponds to a first-order (single photon) resonance, they acquire
a finite lifetime due to scattering into bulk states with the same
quasi-energy. However, these scattering processes involve
high-order resonances (multi-photon processes). For sufficiently weak driving, the lifetime
broadening of the Floquet edge states is parametrically smaller than the size
of the gap in which they reside. Therefore, the chiral Floquet edge
states are still well-defined excitations.

Throughout this work, we have focused on the topological characteristics of the single-particle spectra of periodically-driven two-dimensional systems.
Notably, we have not discussed the filling of these states for fermionic (or bosonic) systems.
This is a very interesting and challenging problem of its own, which is beyond the scope of this work.
However, future studies on the phenomenology of periodically-driven multi-particle systems will need to address this issue in detail.

The invariants presented in this paper give a complete topological description of the Floquet band structures of periodically driven, two-dimensional non-interacting systems, if no additional symmetries are imposed.
In static systems, the presence of additional symmetries leads to a rich topological characterization of non-interacting
systems\cite{Kitaev09,Schnyder08}. In the case of periodically driven systems, some robust topological phenomena have been demonstrated to be stabilized by additional symmetries such as an effective time reversal symmetry\cite{Nate11, Nate3D,Kitagawa11}. A full topological classification for driven systems is yet to be developed, and is an interesting direction for further study.

{\bf Acknowledgements} We thank Gil Refael, Daniel Podolsky, and
P. Zoller for helpful discussions. This work was supported by NSF grants DMR-090647, and PHY-0646094 (M.R.), as well as DMR-1103860, and DMR-0705472 (E.B.). E.B. also acknowledges support from 
the ISF under grant 7113640101.
M.L. acknowledges support from the Alfred P. Sloan Foundation.
N.L. acknowledges support by DARPA under award N66001-12-1-4034.  
N.L. and M.R. acknowledge support provided by the Institute for
Quantum Information and Matter (IQIM), an NSF Physics Frontiers
Center with support of the Gordon and Betty Moore Foundation.  

\appendix

\section{Derivation of the relation between edge modes and winding number} \label{edgewindapp}
In this section, we prove equation (\ref{edgewindform}): we show that the number of Floquet
edge modes is given by the winding number, i.e. $n_{\text{edge}} = W[U]$, for any system in which the
bulk Floquet operator is the identity. Our derivation is inspired by
an analogous derivation in Ref.~\onlinecite{KitaevHoneycomb}. 

The precise statement we will prove is as follows. Let $H$ be a translationally invariant,
periodically driven Hamiltonian  with a time evolution operator $U$. Suppose that $U$ has a trivial Floquet operator -- that is,
$U(\vec{k},T) = \vec{1}$. Now, consider a cylindrical geometry with periodic boundary conditions in the
$y$-direction and open boundary conditions at $x=1$ and $x = L_x$ (here $x$ and $y$ are integers labeling lattice sites).
Consider a Hamiltonian $\tilde{H}$ which is identical to $H$ in the interior of the cylinder,
but can take any form near the boundaries as long as it is local and translationally invariant in the $y$
direction. Let $\tilde{U}$ be the corresponding time evolution operator. What we will show is that
the number of edge modes of $\tilde{U}(\vec{k},T)$ is given by the winding number $W[U]$.

The first step is to find a formal mathematical expression for the number of Floquet edge modes, $n_{\text{edge}}$.
To this end, it is convenient to describe $\tilde{H}$ and $\tilde{U}$ in mixed $x, k_y$ space. In this description,
we denote the Hamiltonian by $\tilde{H}_{xx'}(k_y,t)$ and the time evolution operator by $\tilde{U}_{xx'}(k_y,t)$.
Here we drop the additional indices $\alpha,\alpha'$ corresponding to the
sites within the unit cell, so that $\tilde{H}_{xx'}(k_y,t)$ and $\tilde{U}_{xx'}(k_y,t)$ are actually
$N \times N$ matrices for each $x,x',k_y,t$.

The time evolution operator $\tilde{U}_{xx'}(k_y,t)$ satisfies several important properties.
The first property is that it is quasi-diagonal in $x$: $\tilde{U}_{xx'}(k_y,t) \rightarrow 0$ as
$|x-x'| \rightarrow \infty$. To see this, note that $\tilde{U}_{xx'}$ can be thought
of as a propagator for single particle states. Furthermore, the velocity of these states is
bounded during the course of the time evolution: $|v| \leq c$ for some $c$. It then follows that $\tilde{U}_{xx'}$ vanishes
exponentially if $x,x'$ are separated by a distance greater than $c \cdot t$.

Another important property of $\tilde{U}$ is that the Floquet operator $\tilde{U}_{xx'}(k_y,T)$ reduces to the identity
matrix $\delta_{xx'}$ unless $x$ and $x'$ are near $x=1$ or $x=L_x$ (i.e., within distance $c \cdot T$
of one of the boundaries). This property follows from the same reasoning as above: as long as $x$ and $x'$ are far
from the boundaries,
the amplitude to propagate between them must be identical to that in the translationally invariant system (where
$U(T) = \vec{1}$). Therefore
\begin{equation}
\tilde{U}_{xx'}(k_y,T) = U_{xx'}(k_y, T) =\delta_{xx'},
\end{equation}
unless $x$ and $x'$ are near one of the boundaries.

It is illuminating to write out the matrix $\tilde{U}(k_y, T)$ more explicitly. All together, the matrix
$\tilde{U}(k_y, T)$ has dimension $N L_x \times N L_x$, since the lattice index $x$ runs from $1$ to $L_x$, and the band index $\alpha$ runs from
$1$ to $N$. Let us list the rows and columns in order $x=1,...,L_x$, with the band index running from $1$ to $N$
for each value of $x$. Then, it follows from the above two properties that $\tilde{U}(k_y,T)$ (approximately) takes the block
diagonal form
\begin{equation}
\tilde{U}(k_y,T) = \bpm \tilde{U}_1(k_y) & 0 & 0 \\ 0 & \vec{1} & 0 \\ 0 & 0 & \tilde{U}_{3}(k_y) \epm,
\label{Umatrix}
\end{equation}
where $\tilde{U}_1(k_y)$ and $\tilde{U}_3(k_y)$ are unitary matrices describing the action of $\tilde{U}$ near the
two boundaries and $\vec{1}$ describes the action of $\tilde{U}$ in the interior of the cylinder.
To see this, imagine we partition the lattice sites $x$ into three groups: those with $x \leq L_x/3$,
those with $L_x/3 < x < 2L_x/3$, and finally those with $x \geq 2L_x/3$. In the thermodynamic limit,
$\tilde{U}(k_y,T)$ will be block diagonal with respect to these three groups -- up to corrections which are
exponentially small in $L_x$. Furthermore, $\tilde{U}$ will act like the identity matrix $\vec{1}$ on the
middle group of sites, $L_x/3 < x < 2L_x/3$. Thus, we can define
the three diagonal blocks $\{\tilde{U}_1(k_y),\vec{1},\tilde{U}_3(k_y)\}$ to be $N L_x/3 \times N L_x/3$ matrices
which describe the action of $\tilde{U}$ within these three groups of sites.

Using the above matrix representation for $\tilde{U}$, we can write down a simple expression for the number of edge modes,
$n_{\text{edge}}$, on the $x = L_x$ edge. In particular, we claim that
\begin{equation}
n_{\text{edge}} = -\frac{1}{2\pi i} \int d k_y \ \text{Tr}(\tilde{U}(k_y,T)^{-1} \partial_{k_y} \tilde{U}(k_y,T) \cdot Q),
\label{edgecount}
\end{equation}
where $Q$ is an operator defined by
\begin{equation}
Q_{xx'} = g(x)\delta_{xx'},
\end{equation}
and $g$ is any function satisfying
\begin{equation}
g(x) =
\begin{cases} 0
& \mbox{if $x \leq L_x/3$}  \\
1,  & \mbox{if $x \geq 2L_x/3$}.
\end{cases}
\end{equation}
The trace in Eq.~(\ref{edgecount}) is taken over the band indices $\alpha,\alpha'$ and the site indices $x,x'$.

To understand where (\ref{edgecount}) comes from, imagine we replace $Q$ by the identity operator $\vec{1}$.
Then the above integral reduces to $-\frac{1}{2\pi i} \int dk_y \ \text{Tr}(\tilde{U}^{-1} \partial_{k_y} \tilde{U})$,
which counts the \emph{total} number of chiral modes propagating in the $y$ direction -- both at $x=L_x$ and $x=1$\cite{KitagawaPRB}.
Our claim is that, when we include the operator $Q$ in the expression, we effectively count the edge modes near the
$x=L_x$ boundary, and throw out the contribution from the $x=1$ boundary.

To see this explicitly, let us write out the matrix for $Q$ using the same notation as equation (\ref{Umatrix}):
\begin{equation}
Q = \bpm \vec{0} & 0 & 0 \\ 0 & Q_{2} & 0 \\ 0 & 0 & \vec{1} \epm.
\label{Qmatrix}
\end{equation}
Here $Q_{2}$ describes the action of $Q$ on the sites with $L_x/3 < x < 2L_x/3$ while $\vec{0}$ and $\vec{1}$ describe
the action of $Q$ on the sites with $x \leq L_x/3$ and $x \geq 2L_x/3$, respectively.
From the matrix representations for $Q$ and $\tilde{U}$, we can see that the only contribution to
${\rm Tr}(\tilde{U}^{-1} \partial_{k_y} \tilde{U} \cdot Q)$ comes from the sites near the $x = L_x$ boundary:
\begin{equation}
{\rm Tr}(\tilde{U}^{-1} \partial_{k_y} \tilde{U} \cdot Q) = {\rm Tr}(\tilde{U}_{3}^{-1}(k_y) \partial_{k_y} \tilde{U}_{3}(k_y)).
\end{equation}
Clearly, when we integrate this expression over $k_y$ and take the trace, the result is simply the number of chiral edge modes
localized near the $x = L_x$ boundary, as claimed above.

The next step is to write (\ref{edgecount}) as an integral over $t$:
\begin{equation}
n_{\text{edge}} = -\frac{1}{2\pi i} \int dt dk_y \ \partial_t \text{Tr}( \tilde{U}^{-1} \partial_{k_y} \tilde{U} \cdot Q).
\end{equation}
We then add the total derivative $\partial_{k_y} \text{Tr}(-\tilde{U}^{-1} \partial_t \tilde{U} \cdot Q)$ to the
integrand, thereby deriving
\begin{align}
n_{\text{edge}}
&= \frac{1}{2\pi i} \int dt dk_y \nonumber \\
&\cdot \text{Tr}(\tilde{U}^{-1} \partial_{k_y} \tilde{U} \cdot [Q,\tilde{U}^{-1} \partial_t \tilde{U}]). \nonumber
\end{align}

Next we note that the integrand only receives contributions from the interior of the cylinder, since $Q$ is constant near
the boundaries and hence the commutator $[Q, \tilde{U}^{-1} \partial_t \tilde{U}]$ vanishes there. In the interior
of the cylinder, however, $\tilde{U}$ is identical to $U$. We are therefore free to replace $\tilde{U} \rightarrow U$:
\begin{align}
n_{\text{edge}} &= \frac{1}{2\pi i} \int dt dk_y \nonumber \\
&\cdot \text{Tr}(U^{-1} \partial_{k_y} U \cdot [Q, U^{-1} \partial_t U]).
\label{edgecomm}
\end{align}

To proceed further, let $A \equiv U^{-1} \partial_{k_y} U$, $B \equiv U^{-1} \partial_t U$.
Exploiting the translational invariance of $A,B$, we have
\begin{eqnarray}
\text{Tr}(A \cdot [Q,B]) &=& \sum_{xx'} A_{xx'} B_{x'x} (g(x')-g(x)) \nonumber \\
&=& \sum_{xs} A_{0s}B_{s0} \cdot (g(x+s)-g(x)). \nonumber
\end{eqnarray}
Next, we use the identity $\sum_{x} (g(x+s) - g(x)) = s$ to write this as
\begin{eqnarray}
\text{Tr}(A \cdot [Q,B])
&=& \sum_s A_{0s}B_{s0} \cdot s \nonumber \\
&=& \frac{i}{2\pi} \int dk_x \ A(k_x) \partial_{k_x} B(k_x), \nonumber
\end{eqnarray}
where the last equality comes from taking the Fourier transform.
Substituting this expression into (\ref{edgecomm}), we have
\begin{equation}
n_{\text{edge}} = \frac{1}{4\pi^2} \int dt dk_x dk_y \ \text{Tr}( U^{-1} \partial_{k_y} U \cdot \partial_{k_x} (U^{-1} \partial_t U)).
\nonumber
\end{equation}
The final step is to massage this expression into the desired form (\ref{edgewindform}) by adding a derivative with respect to
$k_x, k_y$, and  $t$ to the integrand. In particular, we add:
\begin{eqnarray}
- \frac{1}{2}\partial_{k_x} \text{Tr}(U^{-1} \partial_{k_y} U \cdot U^{-1} \partial_t U) &+&
\frac{1}{2} \partial_{k_y} \text{Tr}(U^{-1} \partial_{k_x} \partial_t U) \nonumber \\
&-& \frac{1}{2} \partial_t \text{Tr}(U^{-1} \partial_{k_x}\partial_{k_y} U). \nonumber
\end{eqnarray}
Adding this term and simplifying gives
\begin{align}
n_{\text{edge}} &= \frac{1}{8\pi^2} \int dt dk_x dk_y \nonumber \\
&\cdot \text{Tr}(U^{-1} \partial_{k_x} U \cdot U^{-1} \partial_{k_y} U \cdot
U^{-1} \partial_t U) - (k_x \leftrightarrow k_y) \nonumber \\
&= W[U],
\end{align}
as claimed.

\section{Derivation of (\ref{windchernrel})} \label{edgechernapp}
In this section, we derive relation (\ref{windchernrel}) between the winding number $W[U_\varepsilon]$ and the
Chern number of the Floquet bands. Let $U(\vec{k},T)$ be Floquet operator with gaps at $\varepsilon, \varepsilon'$.
We wish to compute the difference in the winding numbers $W[U_{\varepsilon'}] - W[U_\varepsilon]$. A simple way
to do this is to note that $W[U_{\varepsilon'}] - W[U_\varepsilon]$ is equal to the winding number of a map $U$
which ``glues'' $V_{\varepsilon'}$ and $V_\varepsilon$ together with opposite orientations. To be precise,
\begin{equation}
W[U_{\varepsilon'}] - W[U_\varepsilon] = W[U],
\label{nuu}
\end{equation}
where
\begin{equation}
U(\vec{k},t) =
\begin{cases}
V_{\varepsilon}(\vec{k},2t) & \mbox{if } 0 \leq t \leq T/2 \\
V_{\varepsilon'}(\vec{k},2T - 2t) & \mbox{if } T/2 \leq t \leq T.
\end{cases}
\end{equation}
To compute $W[U]$, we use the fact that it is invariant under continuous deformations of $U$.
We note that the above map $U$ can be continuously deformed into $\bar{U}$ where
\begin{equation}
\bar{U}(\vec{k},t) = V_{\varepsilon}(\vec{k},t) \cdot V_{\varepsilon'}(\vec{k},t)^{-1}.
\label{deftW}
\end{equation}
To see this, note that the following interpolation does the job:
\begin{equation}
U_s(\vec{k},t) =
\begin{cases}
V_{\varepsilon}(\vec{k},2t) \cdot V_{\varepsilon'}(\vec{k},2t)^{-s} & \mbox{if } 0 \leq t \leq T/2 \\
V_{\varepsilon'}(\vec{k},2T - 2t)^{1-s} & \mbox{if } T/2 \leq t \leq T,
\end{cases} \nonumber
\end{equation}
for $0 \le s \le 1$.

The problem reduces to computing the winding number $W[\bar{U}]$. Note that $V_\varepsilon$ and $V_{\varepsilon'}$ differ
only in the position of the branch cut of the logarithm in definition (\ref{Vu}), which affects only the quasi-energy
eigenvalues but not the eigenvectors. These two operators can therefore be simultaneously diagonalized, and commute.
From the definition we see that
\begin{equation}
\bar{U}(\vec{k},t) = V_{\varepsilon}(\vec{k},t) \cdot V_{\varepsilon'}(\vec{k},t)^{-1}
= e^{\frac{2\pi i t}{T} P_{\varepsilon \varepsilon'}(\vec{k})},
\end{equation}
where $P_{\varepsilon \varepsilon'}(\vec{k})$ is the projector onto the Floquet eigenstates with crystal momentum $\vec{k}$,
and with eigenvalues between $\varepsilon$ and $\varepsilon'$. Therefore, according to the calculation of Appendix
\ref{windflatapp}, the
corresponding winding number is $W[\bar{U}] = \mathcal{C}_{\varepsilon \varepsilon'}$, where
$\mathcal{C}_{\varepsilon \varepsilon'}$ is the total Chern number of all the Floquet bands with
eigenvalues between $\varepsilon, \varepsilon'$. Putting this all together, we deduce:
\begin{equation}
W[U_{\varepsilon'}] - W[U_\varepsilon] = W[\bar{U}] = \mathcal{C}_{\varepsilon \varepsilon'},
\end{equation}
as we set out to show.

\section{Winding number for time independent, flat band Hamiltonians} \label{windflatapp}
In this section, we compute the value of $W[U]$ for time independent, flat band Hamiltonians of
the form
\begin{equation}
H(\vec{k}) = -\frac{2\pi}{T} P(\vec{k}),
\end{equation}
where $P(\vec{k})$ is an $N \times N$ Hermitian matrix, all of whose eigenvalues
are either $0$ or $1$ (i.e.~$P$ is a projection operator). Note that the prefactor $\frac{2\pi}{T}$ guarantees that the
corresponding Floquet operator is trivial, i.e.~$U(T) = e^{-iHT} = \vec{1}$. For concreteness we assume that $P(\vec{k})$ has $M$ eigenvalues equal
to $1$, and $N-M$ eigenvalues equal to $0$ where $M < N$. Thus, the Hamiltonian $H$ has $M$ bands with
energy $-2\pi/T$ and $N-M$ bands with energy $0$. We will show that $W[U]$ is given by the total Chern number $\mathcal{C}$ of
the bands with energy $-2\pi/T$.

First, we note that the time evolution operator $U = e^{\frac{2\pi i t}{T} P(\vec{k})}$ satisfies the identities
\begin{eqnarray}
U &=& (e^{2\pi i t/T} -1) P + 1 \nonumber \\
U^{-1} &=& (e^{-2\pi i t/T} -1) P + 1,
\end{eqnarray}
from which it follows that
\begin{eqnarray}
\!\!\!\!\!\!\!\!U^{-1} \partial_{k_x} U &=& (e^{-2\pi i t/T} -1)(e^{2\pi i t/T}-1) P \partial_{k_x} P \nonumber \\
&+& (e^{2\pi i t/T}-1) \partial_{k_x} P
\nonumber \\
&\equiv& a \left(\frac{2\pi t}{T} \right) \cdot P \partial_{k_x} P + b \left(\frac{2 \pi t}{T} \right) \cdot \partial_{k_x} P.
\end{eqnarray}
Here,
\begin{align}
a(\theta) = 2 - 2 \cos(\theta) \ , \ b(\theta) = e^{i\theta}-1.
\end{align}
Similarly, we have
\begin{eqnarray}
U^{-1} \partial_{k_y} U &=& a \left(\frac{2\pi t}{T} \right) \cdot P \partial_{k_y} P + b \left(\frac{2\pi t}{T} \right) \cdot \partial_{k_y} P \nonumber \\
U^{-1} \partial_t U &=& \frac{2 \pi i}{T} P.
\end{eqnarray}
Combining these results, we find
\begin{eqnarray}
{\rm Tr}\big([U^{-1} \partial_t U][U^{-1} \partial_{k_x} U &][& U^{-1} \partial_{k_y} U] \big) = \nonumber \\
{\rm Tr}\Big(\frac{2\pi i}{T} P[a \cdot P \partial_{k_x} P  + b \cdot \partial_{k_x} P &][& a \cdot P \partial_{k_y} P
+ b \cdot \partial_{k_y} P]\Big). \nonumber
\end{eqnarray}
Next, we make use of the identity
\begin{eqnarray}
(P \partial_i P) \cdot P &=& (\partial_i(P^2) - \partial_i P P) \cdot P \nonumber \\
&=& \partial_i P \cdot P - \partial_i P \cdot P \nonumber \\
&=& 0 \nonumber
\end{eqnarray}
to reduce the expression to
\begin{eqnarray}
{\rm Tr}([U^{-1} \partial_t U][U^{-1} \partial_{k_x} U &][& U^{-1} \partial_{k_y} U] ) = \nonumber \\
\frac{2\pi i}{T}(a&+&b)b \cdot {\rm Tr}(P\partial_{k_x} P \partial_{k_y} P ). \nonumber
\end{eqnarray}
Substituting in the above expressions for $a,b$, we find
\begin{eqnarray}
{\rm Tr}([U^{-1} \partial_t U][U^{-1} \partial_{k_x} U &][& U^{-1} \partial_{k_y} U]) = \nonumber \\
\frac{2\pi i}{T}(2 \cos(2\pi t/T)-2) &\cdot& {\rm Tr}(P\partial_{k_x} P \partial_{k_y} P).
\end{eqnarray}
We are now ready to compute the winding number corresponding to $U$:
\begin{align}
W[U] &= \frac{1}{8\pi^2} \int dt dk_x dk_y \nonumber \\
&\cdot {\rm Tr}\left([U^{-1} \partial_t U][U^{-1} \partial_{k_x} U] [U^{-1} \partial_{k_y} U] \right)
- (k_x \leftrightarrow k_y) \nonumber \\
&= \frac{1}{8\pi^2} \int dt dk_x dk_y \frac{2\pi i}{T}(2\cos(2\pi t/T)-2) \nonumber \\
&\cdot
{\rm Tr}\left(P \cdot [\partial_{k_x} P ,\partial_{k_y} P ]\right) \nonumber \\
&= \frac{1}{2\pi i} \int dk_x dk_y {\rm Tr}\left(P \cdot [\partial_{k_x} P, \partial_{k_y} P] \right). \label{Wproj}
\end{align}
The latter expression is precisely the Chern number $\mathcal{C}$, Eq.~(\ref{chern2}), of the bands with quasi-energy $\epsilon = -2\pi/T$.

It is also possible to derive this identity from simple physical considerations. To this end, let us calculate the number of Floquet
edge modes $n_{\text{edge}}$ corresponding to $U$ in two different ways. On one hand, we know that $n_{\text{edge}} = W[U]$.
On the other hand, it follows from the special form of $U$ that the number of Floquet edge modes is equal to the number of edge modes of
$H$ with quasi-energies between $-2\pi/T$ and $0$. The latter number is equal to the total Chern number $\mathcal{C}$ of the $\epsilon = -2\pi/T$ bands of $H$. In this way, we deduce that $n_{\text{edge}} = \mathcal{C}$. Combining these two results, the identity (\ref{Wproj}) follows immediately.

\end{document}